\documentstyle[preprint,aps]{revtex}
\begin{document}
\newif\ifplot
\plottrue
\newcommand{\RR}[1]{[#1]}
\newcommand{\intsum}{\sum \kern -15pt \int}
\newfont{\Yfont}{cmti10 scaled 2074}
\newcommand{\Y}{\hbox{{\Yfont y}\phantom.}}
\def\O{{\cal O}}
\newcommand{\bra}[1]{\left< #1 \right| }
\newcommand{\braa}[1]{\left. \left< #1 \right| \right| }
\def\Bra#1#2{{\mbox{\vphantom{$\left< #2 \right|$}}}_{#1}
\kern -2.5pt \left< #2 \right| }
\def\Braa#1#2{{\mbox{\vphantom{$\left< #2 \right|$}}}_{#1}
\kern -2.5pt \left. \left< #2 \right| \right| }
\newcommand{\ket}[1]{\left| #1 \right> }
\newcommand{\kett}[1]{\left| \left| #1 \right> \right.}
\newcommand{\scal}[2]{\left< #1 \left| \mbox{\vphantom{$\left< #1 #2 \right|$}}
\right. #2 \right> }
\def\Scal#1#2#3{{\mbox{\vphantom{$\left<#2#3\right|$}}}_{#1}
{\left< #2 \left| \mbox{\vphantom{$\left<#2#3\right|$}}
\right. #3 \right> }}
\draft
\title{
A New Look into the Partial Wave Decomposition of Three-Nucleon Forces
}
\author{D. H\"uber$^\dagger$,
H. Wita\l a$^*$,
A. Nogga$^\dagger$,
W. Gl\"ockle$^\dagger$,
H. Kamada$^{\ddagger}$
}
\address{
$^\dagger$Institut f\"ur theoretische Physik II, Ruhr-Universit\"at Bochum,
D-44780 Bochum, Germany
}
\address{$^{*}$ Institute of Physics, Jagellonian University, 
PL- 30059 Cracow, Poland}
\address{$^\ddagger$ 
Paul Scherrer Institut, CH-5232, Villigen PSI, Switzerland}
\date{\today}
\maketitle
\widetext
\begin{abstract}
We demonstrate that the partial wave decomposition of three-nucleon forces
used up to now in momentum space has to be necessarily unstable at high
partial waves. This 
does not affect the applications performed up to now, which were restricted to
low partial waves. We present a new way to perform the partial wave
decomposition 
which is free of that defect. This is exemplified for the most often used
$2\pi$-exchange Tuscon-Melbourne three-nucleon force. For the lower partial
waves the results of the old method are reproduced.
\end{abstract}

\pagebreak
\narrowtext

\section{Introduction}

Three-nucleon forces (3NF) act for more than
two nucleons. The interesting question is their strengths and their
signatures. 

A first observable where they clearly show up
is the binding energy of the  triton. Here it is known that the most
recent realistic nucleon-nucleon (NN) forces
cannot produce the triton binding energy and 3NF's are needed in order
to get the experimental number \cite{Sasa} \cite{Pen}.

A next logical step is to look into the 3N continuum. Our results based on
numerically precise solutions of the three-nucleon (3N) Faddeev equations and
realistic NN forces agree overall very well with experimental data
\cite{report}. In elastic Nd scattering there is only one discrepancy, which
clearly sticks out, the low energy vector analyzing powers, which depend
sensitively on the $^{3}P_{j}$ NN force components. There is either an
ambiguity in their determination from NN data or one really sees a 3NF
effect. The inclusion of the 3NF's, however, which have been worked out up to
now, do not diminish that discrepancy \cite{report} \cite{Pisa}. Right now it
remains a puzzle \cite{puzzle}. At very low energies, near the Nd threshold,
one has to expect to see 3NF effects, which are linked to the accompanying
shift of the triton binding energy. But we predict also scattering
observables, which do not scale with the triton binding energy \cite{WB}
\cite{report}. This is an interesting energy region to be studied
experimentally. At higher energies up to about 60 MeV the 3NF effects we find
for cross sections are mostly small (of the order of a
few percent) \cite{report} \cite{Koln} \cite{Krakau} and sometimes even
shifting theory in the 
wrong direction. For certain spin observables there can be larger effects
\cite{report}. We apparently stand at the very beginning in understanding
3NF's. At much higher energies, however, above 100 MeV nucleon lab energy, we
predict big 3NF effects, though still based on a truncated partial wave basis
for the 3NF \cite{report}. 

The present article is meant to provide the theoretical and numerical ground
to add more partial waves to  the 3NF. With the previous algorithm \cite{CG81}
for decomposing the 3NF into partial waves we ran into unexpected numerical
instabilities for higher angular momentum states.

In section \ref{old} we show the origin of this defect using a
model 3NF. We demonstrate that the failure in the 
partial wave decomposition (PWD) used up to now cannot be cured for higher
angular momenta.

In section \ref{new} we introduce a new PWD. Thereby we decompose
 the 3NF into two quasi
two-body parts and a recoupling of Jacobi variables in between.
This leads to a new PWD, which is free of the inherent difficulties of the old
PWD. We perform the new PWD for the $2\pi$-exchange
Tucson-Melbourne (TM) 3NF \cite{Coon} \cite{CG81}. Thereby we also comment on
the numerics where it is relevant.

We demonstrate in a numerical study in section \ref{comp} that
for lower partial 
waves the old and new PWD give the same results for Nd scattering. In a bound
state calculation we extend the new PWD to higher partial waves and reproduce
exactly 
the result of the Los Alamos - Iowa group \cite{LA}.

Finally we summarize in section \ref{sum}.

Appendix \ref{perm} contains
the expressions for the permutation operators which are needed in
our new approach. They are
generalizations of the permutation operators which we usually use \cite{HWG93}
\cite{Buch} in the Faddeev scheme. 

The permutation operators lead to
interpolation of very many matrix elements. This includes a two-fold
interpolation, as is shown in appendix \ref{perm}, for which a very efficient
and accurate algorithm has to be 
used. It turned out, that our usual scheme, namely cubic basis
splines\cite{GHN82}, requires too much CPU time, which would not allow to
calculate  
the 3NF within a reasonable time. Therefore we use cubic hermitian splines
which 
need much less operations for one interpolation than the basis splines. In
appendix \ref{spline} we briefly review the introduction of cubic hermitian
splines and rewrite them to the most efficient form for our purposes.

\section{The Pitfall in the Old Partial Wave Decomposition}
\label{old}

\begin{figure}
\input{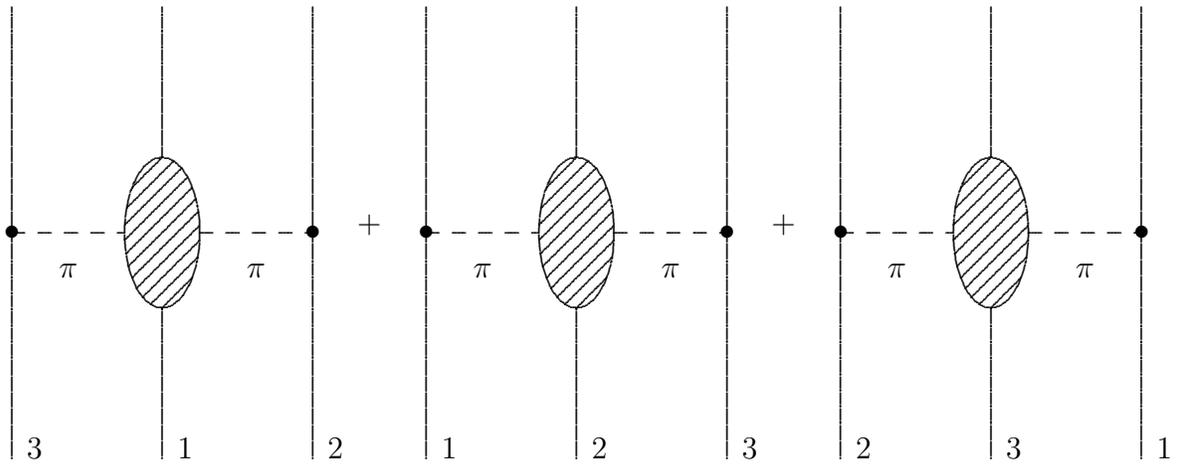}
\caption{The $2\pi$-exchange 3NF.}
\label{V4}
\end{figure}

The $2\pi$-exchange 3NF is shown in Fig.~\ref{V4}. The three diagrams
in Fig.~\ref{V4} differ only by cyclical and anti-cyclical permutations
of the three particles. Therefore it is sufficient to calculate the first
diagram and find the 
full 3NF by the application of proper permutation 
operators. The first diagram is denoted by $V_4^{(1)}$ since particle 1 is
singled out. 

We need the matrix elements of $V_4^{(1)}$ in a basis
$\ket{pq\alpha}$, where $p$ and $q$ are Jacobi momenta ($\vec p$ is the
relative 
momentum between particles 2 and 3 and $\vec q$ is the momentum of particle 1
relative to the pair 2-3) and $\alpha$ denotes discrete quantum numbers. In
detail 
\begin{equation}
\label{1}
\ket{pq\alpha}\equiv \ket{p(ls)j\ q(\lambda {1\over 2})I\ (jI)JM;\ 
(t{1\over 2})TM_{T}}
\end{equation}
where 
$l$, $s$ and $j$ are orbital angular momentum, spin and total angular momentum in the
two-body subsystem 2-3. $\lambda$, $1\over 2$ and $I$ are the corresponding
quantities 
for particle 1, and $JM$ denote the conserved total 3N angular
momentum and its magnetic quantum number. In iso-spin space $t$ refers to
particles 2 and 3, which couples with $1/2$ to total iso-spin $T$ and
its magnetic quantum number $M_{T}$. 
Thus the matrix elements we are looking for are of the form
$\bra{pq\alpha}V_4^{(1)}\ket{p'q'\alpha'}$. 

We found that for $p=p'$ large, $l=l'\ge 4$, $q=q'$ large and 
$\lambda=\lambda' \ge 4$ the matrix 
elements $\bra{pq\alpha }V_4^{(1)}\ket{p'q'\alpha'}$
are some orders of magnitude larger than neighbouring non symmetric
matrix elements. The increase of $p=p'$, $q=q'$, $l=l'$ or $\lambda=\lambda'$ 
leads to an
explosion in magnitude of the matrix elements. Moreover, we found that in
those 
partial waves the symmetry of $\bra{pq\alpha
  }V_4^{(1)}\ket{p'q'\alpha'}$ under exchange of the primed and unprimed
variables, which 
is manifest by the operator structure, is violated. This symmetry violation
happens even for smaller values of $l$ or $\lambda$.

Such a behaviour hints clearly to a numerical instability of the
code. This instability is not caused by an error in the code nor in the formal
expressions of the PWD, but by the numerical realisation on a computer which
always relies on a finite number of digits. We shall now illustrate the
problem for a model using the old PWD.

The instability is connected to the values of the momenta 
$p$, $p'$, $q$ and $q'$ and the corresponding angular momentum quantum numbers
$l$, $l'$, $\lambda$ and $\lambda'$ and is not related to spin or iso-spin.
Thus we can use a spin and iso-spin independent 3NF model for our purpose.
Moreover, the behaviour in $q$ and $\lambda$ turns out to be analogical to
$p$ and $l$ and therefore it suffices to treat only the $p$ and $l$ dependence
in the 3NF model.
(The old PWD for the full TM 3NF, which we used up to now, 
can be found in \cite{CG81} and in more detail in \cite{PhD}.)

Let us consider $V_4^{{(1)}}$ as depicted in
Fig.~\ref{V4(1)}. Neglecting spin and iso-spin, meson-nucleon form factors and
replacing the shaded blop by a constant, we
are
simply left with the two pion propagators
\begin{equation}
\label{2}
V_{4}^{(1)}\Rightarrow {1\over {\vec Q}^{2}+m_{\pi}^{2}}\ 
            {1\over {\vec {Q'}}^{2}+m_{\pi}^{2}}
\end{equation}
where $\vec Q$ and $\vec {Q'}$ are given by
\begin{eqnarray}
\label{3}
\nonumber
\vec Q &=& \vec p - \vec {p'} - {1\over 2} (\vec q -\vec {q'})\\
\vec {Q'} &=& \vec p - \vec {p'} + {1\over 2} (\vec q -\vec {q'})
\end{eqnarray}

We further simplify by freezing
 the $\vec q$-dependence and taking only one
of the two propagators.
Since the
problems occur for $p=p'$ we choose $|\vec p|=|\vec {p'}|$ and allow only for
an angular variation. Thus the model 
3NF becomes
\begin{equation}
\label{4}
V_{4}^{model}={1\over (p(\hat p-\hat {p'})+\vec a)^{2}+m_{\pi}^{2}}
\end{equation}
where $\vec a$ is a fixed vector.

We define the two angles
\begin{equation}
\label{5}
x_{2}\equiv \hat p \cdot \hat {p'}\ ;\ \ \ 
x_{1}\equiv (\widehat{\vec p -\vec {p'}})\cdot
\hat a
\end{equation}
and rewrite eq.~(\ref{4}) as
\begin{eqnarray}
\label{6}
\nonumber
V_{4}^{model}&=&
{1\over 2p^{2}(1-x_{2})+a^{2}+2ap\sqrt{2(1-x_{2})}x_{1}+m_{\pi}^{2}}\\
&\equiv & {1\over B+Ax_{1}}={A^{-1}\over {B\over A}+x_{1}}
\end{eqnarray}
with
\begin{eqnarray}
\label{7}
\nonumber
A&=&2ap\sqrt{2(1-x_{2})}\\
B&=&2p^{2}(1-x_{2})+a^{2}+m_{\pi}^{2}
\end{eqnarray}
The expression in eq.~(\ref{6}) is expanded in Legendre polynomials resulting
in 
\begin{equation}
\label{8}
V_{4}^{model}=A^{-1}\sum_{l_{1}=0}^{\infty}\hat l_{1}\
P_{l_{1}}(x_{1})\
Q_{l_{1}}\left({B\over A}\right)
\end{equation}
Here and in the following we use the abbreviation
\begin{equation}
\label{8a}
\hat l\equiv 2l+1
\end{equation}
for orbital and spin angular momenta.

The Legendre polynomials themselves can be rewritten in terms of coupled
spherical harmonics with respect to the two directions involved in $x_{1}$:
\begin{equation}
\label{9}
P_{l_{1}}((\widehat{\vec p -\vec {p'}})\cdot \hat a)=
{4\pi \over \sqrt{\hat l_{1}}}\ (-)^{l_{1}}\ 
\Y_{l_{1}l_{1}}^{00}(\widehat{\vec p -\vec {p'}},\hat a)
\end{equation}
with
\begin{equation}
\label{10}
\Y_{l_{1}l_{2}}^{lm}(\hat a,\hat b)\equiv 
\sum_{\nu} C(l_{1}\nu \ l_{2}m-\nu ,lm)\
Y^{\nu}_{l_{1}}(\hat a)\ Y^{m-\nu}_{l_{2}}(\hat b)
\end{equation}
Further one has
\begin{equation}
\label{11}
Y^{m}_{l}(\widehat{\vec a+\vec b})=\sum_{\lambda_{1}+\lambda_{2}=l}
{a^{\lambda_{1}}b^{\lambda_{2}}\over |\vec a+\vec b|^{l}}\
\sqrt{4\pi \ \hat l!\over \hat \lambda_{1}!\ \hat \lambda_{2}!}\
\Y_{\lambda_{1}\lambda_{2}}^{lm}(\hat a,\hat b)
\end{equation}

We insert now eq.~(\ref{9}) and eq.~(\ref{11}) into eq.~(\ref{8}) and get
\begin{eqnarray}
\label{12}
\nonumber
V_{4}^{model}&=&A^{-1}\sum_{l_{1}=0}^{\infty}\hat l_{1}\
Q_{l_{1}}\left({B\over A}\right)
{4\pi \over \sqrt{\hat l_{1}}}\ (-)^{l_{1}}\\
&\times&\sum_{\lambda_{1}+\lambda_{2}=l_{1}}
{p^{\lambda_{1}}(-{p'})^{\lambda_{2}}\over |\vec p-\vec {p'}|^{l_{1}}}\
\sqrt{4\pi \ \hat l_{1}!\over \hat \lambda_{1}!\ \hat \lambda_{2}!}\
\left\{ \Y_{\lambda_{1}\lambda_{2}}^{l_{1}m}(\hat p,\hat {p'}),
Y_{l_{1}}(\hat a)\right\}^{00}
\end{eqnarray}
Hereby we used the notation
\begin{equation}
\label{12a}
\{ a_{l_{1}},b_{l_{2}}\}^{lm}\equiv 
\sum_{\nu} C(l_{1}\nu \ l_{2}m-\nu ,lm)\
a^{\nu}_{l_{1}}\ b^{m-\nu}_{l_{2}}
\end{equation}
to describe the coupling of two quantities $a$ and $b$ with angular momenta
$l_{1}$ and $l_{2}$ to $lm$.
Already here we can see the problem with this type of PWD. The expression in
eq.~(\ref{12}) seems to have a 
singularity for 
$\vec p=\vec {p'}$, though the 
model force is not singular, of course. The singularity in
eq.~(\ref{12}) is only apparent and is removed by the sum over
$\lambda_{1}+\lambda_{2}$, if that sum can be evaluated accurately. The next
step in the PWD is to expand the 
$x_{2}$-dependence in terms of Legendre polynomials analogically to what we
have 
done for the $x_{1}$-dependence. The $x_{2}$ dependence occurs in the
denominator $|\vec p-\vec {p'}|^{l_{1}}$ and in the Legendre polynomial of the
second kind, $Q_{l}(B/A)$. Performing that expansion we separate the singular
denominator from the sum over $\lambda_{1}+\lambda_{2}$, which has to cancel
the
singularity. This separation and the necessarily limited accuracy in the
numerical evaluation of the sum over complicated geometrical expressions is
the reason for the numerical instability as we shall demonstrate now.

Let us now perform the $x_{2}$ expansion:
\begin{eqnarray}
\label{13}
\nonumber
V_{4}^{model}&=&{1\over 2ap}
\sum_{l_{1}=0}^{\infty}4\pi \sqrt{\hat l_{1}} (-)^{l_{1}}
\sum_{\lambda_{1}+\lambda_{2}=l_{1}} (-)^{\lambda_{2}}
\sqrt{4\pi\ \hat l_{1}!\over \hat \lambda_{1}!\ \hat \lambda_{2}!}\
\left\{ \Y_{\lambda_{1}\lambda_{2}}^{l_{1}}(\hat p,\hat {p'}),
Y_{l_{2}}(\hat a)\right\}^{0}\\
&\times& \sum_{l_{2}=0}^{\infty} 2\pi\sqrt{\hat l_{2}}(-)^{{l_{2}}}\
\Y_{l_{2}l_{2}}^{00}(\hat p,\hat {p'})\ H_{l_{1}l_{2}}(p,a)
\end{eqnarray}
The expansion coefficients $H_{l_{1}l_{2}}(p,a)$ are given by
\begin{equation}
\label{14}
H_{l_{1}l_{2}}(p,a)=\int_{-1}^{+1}dx_{2}\ P_{l_{2}}(x_{2})\
{Q_{l_{1}}\left( {2p^{2}(1-x_{2})+a^{2}+m_{\pi}^{2}
\over 2ap\sqrt{2(1-x_{2})}}\right)
\over \left( \sqrt{2(1-x_{2})}\right)^{l_{1}+1} }
\end{equation}

The rest of the PWD is straightforward: one has to recouple the spherical
harmonics in eq.~(\ref{13}) in such a way, that $Y_{\lambda_{1}}(\hat p)$
gets coupled with $Y_{\l_{2}}(\hat p)$ and $Y_{\lambda_{2}}(\hat {p'})$
with $Y_{\l_{2}}(\hat {p'})$. The intermediate form can be simplified using 
\begin{equation}
\label{15}
\Y_{l_{1}l_{2}}^{lm}(\hat a,\hat a)=
\sqrt{\hat l_{1}\ \hat l_{2}\over 4\pi\ \hat l}\
C(l_{1}0\ l_{2}0,l0)\ Y_{l}^{m}(\hat a)
\end{equation}
 Finally projecting onto definite orbital angular momenta one ends up with
\begin{eqnarray}
\label{16}
\nonumber
\bra{plm}V_{4}^{model}\ket{p'l'm'}&=&\pi\ {(-)^{m'}\over ap}
\sum_{l_{1}} \sqrt{\hat l_{1}}
\sum_{l_{2}} \sqrt{\hat l_{2}}\
H_{l_{1}l_{2}}(p,a)\ Y_{l_{1}}^{-m'}(\hat a)\\
\nonumber
&\times& \sum_{\lambda_{1}+\lambda_{2}=l_{1}} 
\sqrt{4\pi\ \hat l_{1!}\over \hat \lambda_{1}!\ \hat \lambda_{2}!}\
\sqrt{\hat l_{2}\hat \lambda_{1}\hat \lambda_{2}}\
\left\{ \matrix{l_{1}&\lambda_{1}&\lambda_{2}\cr l_{2}&l'&l}\right\} \\
\nonumber
&\times& C(l_{2}0\ \lambda_{1}0,l0)\ C(l_{2}0\ \lambda_{2}0,l'0)\\
&\times& C(l_{1}m'\ l_{1}-m',00)\ C(lm\ l'm'-m,l_{1} m')
\end{eqnarray}

We see from eq.~(\ref{16}) that the $p$-dependence is now shifted into the
expansion coefficients $H_{l_{1}l_{2}}(p,a)$. Since the code for the 3NF
failed for large $p$ we regard the limes of
$H_{l_{1}l_{2}}(p,a)$ for $p\to\infty$. 
Inspection of eq.~(\ref{14}) suggests to introduce as a new integration
variable the argument of $Q_{l_{1}}$:
\begin{equation}
\label{17}
z(x_{2})\equiv {2p^{2}(1-x_{2})+a^{2}+m_{\pi}^{2}
\over 2ap\sqrt{2(1-x_{2})}}
\end{equation}
Because $z(x_{2})$ has a minimum at $x_{2}=1-{a^{2}+m_{\pi}^{2}\over 2p^{2}}$
the integral over $z$ splits into two parts.
It results
\begin{eqnarray}
\label{20}
\nonumber
H_{l_{1}l_{2}}(p,a)&=&
\int_{4p^{2}+a^{2}+m_{\pi}^{2}\over 4ap}^{\sqrt{a^{2}+m_{\pi}^{2}}\over a} dz\
\left( -{2a^{2}z\over p^{2}}- {a\over p^{2}}\
{2a^{2}z^{2}-(a^{2}+m_{\pi}^{2})\over
\sqrt{a^{2}z^{2}-(a^{2}+m_{\pi}^{2})}}\right) \\
\nonumber
&\times& P_{l_{2}}\left( 1+{a^{2}+m_{\pi}^{2}\over 2p^{2}}-{a^{2}z^{2}\over p^{2}}
- {az\over p^{2}}\ \sqrt{a^{2}z^{2}-(a^{2}+m_{\pi}^{2})}\right) \\
\nonumber
&\times& Q_{l_{1}}(z)\ \left( 2\left( -{a^{2}+m_{\pi}^{2}\over 2p^{2}}
+ {a^{2}z^{2}\over p^{2}}
+ {az\over p^{2}}\ \sqrt{a^{2}z^{2}-(a^{2}+m_{\pi}^{2})}\right)\right) 
^{-{l_{1}+1\over 2}}\\
&+&\int^{\infty}_{\sqrt{a^{2}+m_{\pi}^{2}}\over a} dz\
\left( -{2a^{2}z\over p^{2}}+ {a\over p^{2}}\
{2a^{2}z^{2}-(a^{2}+m_{\pi}^{2})\over
\sqrt{a^{2}z^{2}-(a^{2}+m_{\pi}^{2})}}\right) \\
\nonumber
&\times& P_{l_{2}}\left( 1+{a^{2}+m_{\pi}^{2}\over 2p^{2}}-{a^{2}z^{2}\over p^{2}}
+ {az\over p^{2}}\ \sqrt{a^{2}z^{2}-(a^{2}+m_{\pi}^{2})}\right) \\
\nonumber
&\times& Q_{l_{1}}(z)\ \left( 2\left( -{a^{2}+m_{\pi}^{2}\over 2p^{2}}
+ {a^{2}z^{2}\over p^{2}}
- {az\over p^{2}}\ \sqrt{a^{2}z^{2}-(a^{2}+m_{\pi}^{2})}\right)\right) 
^{-{l_{1}+1\over 2}}\\
\end{eqnarray}
Because $Q_{l_{1}}(z)$ decreases like $z^{-l_{1}-1}$ for large $z$, 
$z$ is limited to small
values and both arguments of $P_{l_{2}}$ go to 1 for $p\to \infty$. As a
consequence 
$H_{l_{1}l_{2}}(p,a)$ becomes independent of $l_{2}$. We end up with
\begin{equation}
\label{21}
\lim_{p\to \infty }H_{l_{1}l_{2}}(p,a)\propto p^{l_{1}-1}
\end{equation}

Because the model force decreases for large momenta the increase of $H$ for
$p\to 
\infty$ has to be canceled by the sums occurring in
eq.~(\ref{16}). Indeed this is the case, as we will show now.

If $H$ is independent of $l_{2}$ we can perform the sum over $l_{2}$
analytically:
\begin{eqnarray}
\label{23}
\nonumber
&\phantom =&\sum_{l_{2}}\hat l_{2}\
\left\{ \matrix{l_{1}&\lambda_{1}&\lambda_{2}\cr l_{2}&l'&l}\right\}\
C(l_{2}0\ \lambda_{1}0,l0)\ C(l_{2}0\ \lambda_{2}0,l'0)\\
&=&(-)^{\lambda_{2}}\ \sqrt{\hat {l'}\over \hat l_{1}}\
C(\lambda_{1}0\ \lambda_{2}0,l_{1}0)\ C(l_{1}0\ l'0,l0)
\end{eqnarray}
Using this result we can also perform the sum over $\lambda_{1}+\lambda_{2}$
analytically:
\begin{eqnarray}
\label{24}
\nonumber
&\phantom =&\sum_{\lambda_{1}+\lambda_{2}=l_{1}}(-)^{\lambda_{2}}\
\sqrt{4\pi\ (2l_{1})!\over (2\lambda_{1})!\ (2\lambda_{2})!}\
C(\lambda_{1}0\ \lambda_{2}0,l_{1}0)\\
\nonumber
&=&\sum_{\lambda_{1}+\lambda_{2}=l_{1}}(-)^{\lambda_{2}}\
\sqrt{4\pi\ (2l_{1})!\over (2\lambda_{1})!\ (2\lambda_{2})!}\
\sqrt{\hat l_{1}}\ 
\sqrt{(2l_{1}-2\lambda_{1})!\ (2l_{1}-2\lambda_{2})!\over (2l_{1}+1)!}\
{l_{1}!\over (l_{1}-\lambda_{1})!\ (l_{1}-\lambda_{2})!}\\
\nonumber
&=&(-)^{l_{1}}\ \sqrt{4\pi}\ \sum_{\lambda_{1}}(-)^{\lambda_{1}}\
{l_{1}(l_{1}-1)\cdots (l_{1}-\lambda_{1})\over \lambda_{1}!}\\
&=&(-)^{l_{1}}\ \sqrt{4\pi}\ \sum_{\lambda_{1}}(-)^{\lambda_{1}}\
\left( \matrix{l_{1}\cr \lambda_{1}}\right)
=(-)^{l_{1}}\ \sqrt{4\pi}\ (1-1)^{l_{1}}
\end{eqnarray}
From this result we can see two things: firstly the term $(1-1)^{l_{1}}$
cancels indeed 
$p^{l_{1}-1}$. (A somewhat harder inspection shows that only $p^{-2}$
survives, as it should.)
 But secondly, to
achieve this cancellation one has to add up an alternating sum. This causes
loss of digits if the consecutive numbers are of similar magnitude. This is
the case here.
 For large $l_{1}$ it
will be impossible to get a reasonable result for $(1-1)^{l_{1}}$.

Looking into the corresponding sums in the code for the full TM 3NF we found,
that for diagonal matrix elements with $p\ge 6$ fm$^{-1}$, $l=4$, $q\ge 12$
fm$^{-1}$ and $\lambda \ge 4$ this cancellation occurs with consecutive numbers
being of alternating sign and equal within up to 8 digits. 
 Thereby we lost all numerical accuracy. As a consequence the increase
 $p^{l_{1}-1}$ from the corresponding $H$ could not be compensated and badly
 wrong  results appeared. 

It is also clear that an increase of the numerical accuracy cannot solve that
problem. Higher numerical accuracy would only shift the problem to higher
$l$ and $\lambda$. 

\section{The New Partial Wave Decomposition}
\label{new}

\begin{figure}
\input{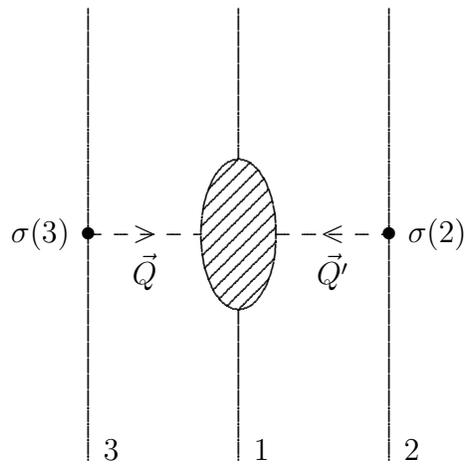}
\caption{The part $V_4^{(1)}$ of the $2\pi$-exchange 3NF.}
\label{V4(1)}
\end{figure}

In this section we will present a new PWD, which avoids the introduction of
artificial singularities.
The new PWD
leads to a numerically stable code. We perform this PWD for
the TM $2\pi$-exchange 3NF. In the notation from Fig.~\ref{V4(1)}, this force
is given by \cite{CG81} 
\begin{eqnarray}
\label{24a}
\nonumber
V_{4}^{(1)}&=&{1\over (2\pi)^{6}}\ {g^{2}\over 4m_{n}^{2}}\
\vec \sigma_{2}\cdot \vec {Q'}\ \vec \sigma_{3}\cdot \vec {Q}\
{F^{2}({\vec {Q}}^{2})\over {\vec {Q}}^{2}+m_{\pi}^{2}}\
{F^{2}({\vec {Q'}}^{2})\over {\vec {Q'}}^{2}+m_{\pi}^{2}}\\
&\times& \left[ \vec \tau_{2}\cdot \vec \tau_{3}\ 
\left( a+b\ \vec Q\cdot \vec {Q'}+c\ ({\vec {Q}}^{2}+{\vec {Q'}}^{2})\right)
+d\ \vec \tau_{1}\cdot \vec \tau_{2}\times\vec \tau_{3}\ 
\sigma_{1}\cdot \vec {Q}\times \vec {Q'} \right]
\end{eqnarray}
The constants
$a$, $b$, $c$ and $d$ are determined by low energy theorems \cite{Coon}.
The form factors are parametrized as
\begin{equation}
\label{25}
F({\vec Q}^{2})\equiv {\Lambda^{2}-m_{\pi}^{2}\over \Lambda^{2}+{\vec Q}^{2}}
\end{equation}
with the cut-off parameter $\Lambda$.

The Faddeev scheme \cite {Buch} \cite{HWG93} \cite{Stadler} \cite{Newal} we
are using requires matrix 
elements \hfill\break
$\bra {pq\alpha} V_{4}^{(1)}\ (1+P) \ket{p'q'\alpha'}$, where $P$ is the sum
of a cyclical and an anti-cyclical permutation of the three
nucleons. Henceforth we  
add a subscript 1, 2 or 3 to the basis states $\ket{pq\alpha}$ in order to
distinguish the three different choices of Jacobi momenta, which single out
either particles 1, 2 or 3 to be the ``spectator nucleon''. In that notation we
can write 
\begin{eqnarray}
\label{29d}
\nonumber
&&\Bra 1{pq\alpha }V_{4}^{(1)}\ (1+P)\ket{p'q'\alpha'}_{1}\\
&=&\intsum''\
\Bra 1{pq\alpha }V_{4}^{(1)}\ket{p''q''\alpha''}_{3}\
\Bra 3{p''q''\alpha''}(1+P)\ket{p'q'\alpha'}_{1}
\end{eqnarray}
We introduced intermediate states of the type 3, since they single out the
pair $(12)$, which participates in one of the two pion exchanges in the
operator $V_{4}^{(1)}$. This is the natural basis to evaluate that specific
pion exchange. The other pion exchange is in the pair $(13)$, for which the
basis $\ket {pq\alpha}_{2}$ is the natural one. This leads us to the basis
idea of the new PWD, to split each term of $V_{4}^{(1)}$ into two parts, which
are related to the separate one pion exchanges and which are then necessarily
linked by a change of coordinates. We call the corresponding two parts
$v_{4}^{(1)}|_{3}$  and $v_{4}^{(1)}|_{2}$, where the subscripts stand again
for the spectator nucleon. Thus each of the four parts of $V_{4}^{(1)}$ in
eq.~(\ref{24a}), $v_{4}^{(1)}$ called in the following, can be cast into the
form 
\begin{eqnarray}
\label{26}
\nonumber
&&\Bra 1{pq\alpha }v_{4}^{(1)}\ket{p'q'\alpha'}_{3}
=\intsum'' \intsum''' \intsum'''' \
\Scal 1{pq\alpha }{p''q''\alpha''}_{2}\\
\nonumber
&\times&
\Bra 2{p''q''\alpha''}v_{4}^{(1)}|_{2}\ket{p'''q'''\alpha'''}_{2}\
\Scal 2{p'''q'''\alpha'''}{p''''q''''\alpha''''}_{3}\\
&\times&
\Bra 3{p''''q''''\alpha''''}v_{4}^{(1)}|_{3}\ket{p'q'\alpha'}_{3}
\end{eqnarray}
The advantage of this PWD is that the matrix elements
$\Bra 2{p''q''\alpha''}v_{4}^{(1)}|_{2}\ket{p'''q'''\alpha'''}_{2}$ and 
$\Bra
3{p''''q''''\alpha''''}v_{4}^{(1)}|_{3}\ket{p'q'\alpha'}_{3}$ are 
easily evaluated, and neither these matrix elements nor the two changes of
basis states introduce numerical instabilities.
These changes of basis states are performed by the
application of certain permutation operators, which are given in appendix
\ref{perm}. Their actual use requires 
2-fold interpolations which are performed with the help of
cubic hermitian splines (see appendix \ref{spline}). 

The basis states $\ket {pq\alpha}_{1}$ in eq.~(\ref{29d}) are antisymmetric in
the 
two-body subsystem $(23)$. This is required in the context of the Faddeev
equations \cite{Buch} \cite{HWG93} \cite{Stadler} \cite{Newal}. The states
$\ket {pq\alpha}_{2}$ 
occurring in 
eq.~(\ref{26}) do not have that property, since the permutation operators
remove that symmetry and therefore one has to sum without restriction over all
quantum numbers to a given total angular momentum $J$ and parity. 
On the other hand the states of type 3 have again to be chosen antisymmetrical
in the 
corresponding two-body subsystem $(12)$. This follows from
eq.~(\ref{29d}). The permutation operator $(1+P)$ applied onto $\ket
{p'q'\alpha'}_{1}$ in eq.~(\ref{29d}) generates a totally antisymmetrical state
and consequently the intermediate states $\Bra 3{p''q''\alpha''}$ have to be
antisymmetrical in the subsystem $(12)$. The
expression (\ref{29d}) can further be simplified. One has
\begin{equation}
\label{26a}
P_{13}\ P_{23} \ket {pq\alpha}_{1}=\ket {pq\alpha}_{3}
\end{equation}
and
\begin{equation}
\label{26b}
P_{23}\ P_{13}\ (1+P) \equiv P_{23}\ P_{13}\ (1+P_{12}\ P_{23}+P_{13}\ P_{23})
=1+P
\end{equation}
Consequently
\begin{eqnarray}
\label{29e}
\nonumber
&&\Bra 1{pq\alpha }V_{4}^{(1)}\ (1+P)\ket{p'q'\alpha'}_{1}\\
&=&\intsum''\
\Bra 1{pq\alpha }V_{4}^{(1)}\ket{p''q''\alpha''}_{3}\
\Bra 1{p''q''\alpha''}(1+P)\ket{p'q'\alpha'}_{1}
\end{eqnarray}

The basis states defined in eq.~(\ref{1}) are a direct product of space-spin
states and iso-spin states: 
\begin{equation}
\label{27}
\ket{pq\alpha } \equiv \ket{pq\alpha_{J}}\ket{\alpha_{T}}
\equiv\ket{p(ls)j\ q(\lambda {1\over 2})I\ (jI)JM}\ket{(t{1\over 2})TM_{T}}
\end{equation}
The iso-spin dependence of $v_{4}^{(1)}$ factorizes off and one is
therefore naturally lead not to include the iso-spin dependence into the
splitting of $v_{4}^{(1)}$ and to evaluate the iso-spin matrix elements $\Bra
2{\alpha_{T}''} \ldots \ket {\alpha_{T}'}_{3}$ directly. One easily finds
\begin{equation}
\label{28}
\Bra 2{(t{1\over 2})TM_{T}}\vec \tau_{2}\cdot \vec \tau_{3}
\ket{(t'{1\over 2})T'M_{T}'}_{3}
=\delta_{TT'}\ \delta_{M_{T}M_{T'}}\ (-6)\ (-)^{t}\ \sqrt{\hat t\hat {t'}}\
\left\{ \matrix{{1\over 2}&{1\over 2}&t'\cr
                {1\over 2}&1&{1\over 2}\cr
                t&{1\over 2}&T}\right\}
\end{equation}
and
\begin{eqnarray}
\label{29}
\nonumber
&&\Bra 2{(t{1\over 2})TM_{T}}\vec \tau_{1}\cdot 
\vec \tau_{2}\times \vec \tau_{3}\ket{(t'{1\over 2})T'M_{T}'}_{3}\\
&&=\delta_{TT'}\ \delta_{M_{T}M_{T'}}\ 24i\ (-)^{2T}\ \sqrt{\hat t\hat {t'}}\
\sum_{\lambda={1\over 2}}^{t+{1\over 2}} (-)^{3\lambda+{1\over 2}}\
\left\{ \matrix{\lambda&{1\over 2}&1\cr
                {1\over 2}&{1\over 2}&t}\right\} \
\left\{ \matrix{T&{1\over 2}&t\cr
                {1\over 2}&1&\lambda \cr
                t'&{1\over 2}&{1\over 2}}\right\}
\end{eqnarray}

These matrix elements will be denoted by $\Bra
2{\alpha_{T}}I_{4}^{(1)}\ket{\alpha_{T}'}_{3}$ in the following. Restricting
the splitting of $v_{4}^{(1)}$ then to the space-spin parts and calling the
two parts $w_{4}^{(1)}|_{2}$ and $w_{4}^{(1)}|_{3}$ eq.~(\ref{26}) achieves
the form 
\begin{eqnarray}
\label{29a}
\nonumber
&&\Bra 1{pq\alpha }v_{4}^{(1)}\ket{p'q'\alpha'}_{3}
=\intsum'' \intsum^{1} \intsum^{2}\
\Scal 1{pq\alpha }{p''q''\alpha''}_{2}\\
\nonumber
&\times&
\left(
\Bra 2{p''q''\alpha_{J}''}w_{4}^{(1)}|_{2}\ket{p_{1}q_{1}\alpha_{1J}}_{2}\
\Scal 2{p_{1}q_{1}\alpha_{1J}}{p_{2}q_{2}\alpha_{2J}}_{3}
\right. \\
\nonumber
&\times&
\left.
\Bra 3{p_{2}q_{2}\alpha_{2J}}w_{4}^{(1)}|_{3}\ket{p'q'\alpha_{J}'}_{3}\
\right) \\
&\times&
\Bra 2{\alpha_{T}''}I_{4}^{(1)}\ket{\alpha_{T}'}_{3}
\end{eqnarray}
In eq.~(\ref{29a}) we introduced intermediate sums $\displaystyle \intsum^{1}$
and $\displaystyle \intsum^{2}$ where the related quatities carry the indices
1 and 2. 
Since the quasi two-body forces $w_{4}^{(1)}|_{2,3}$ are diagonal in the
spectator quantum numbers, the intermediate sums over $\alpha_{1J}$ and
$\alpha_{2J}$ are quite restricted and the main introduction of auxiliary
states in eq.~(\ref{29a}) are the ones with $\alpha''$, which moreover are
unconstrained by any 
symmetry. The $\alpha'$ states are anti-symmetrical as mentioned above but
otherwise unconstrained due to eq.~(\ref{29e}).
We should also note that the operators $w_{4}^{(1)}|_{2,3}$ can
change the parity and the total angular momenta by $\pm 1$, as will be shown
below. 

The evaluation of the permutation operators in eqs.~(\ref{29e}) and
(\ref{29a}) 
consumes essentially all the CPU time; the time for the calculation of the
$w_{4}^{(1)}|_{2,3}$ matrix elements is negligible.

Let us now regard the matrix elements for $w_{4}^{(1)}|_{2,3}$ related to
the four terms proportional to the constants $a$, $b$, $c$ and $d$ in
eq.~(\ref{24a}). 
We omit the common factor ${1\over (2\pi )^{6}}\ {g^{2}\over
  4m_{N}^{2}}$ and the constants $a$, $b$, $c$ and $d$.

The $a$-term naturally splits into two
parts, one related to a pion exchange between particles 1 and 3,
\begin{equation}
\label{30a}
w_{4}^{(1)}|_{2}^{a}=F^{2}(\vec Q^{2})\ 
{\vec \sigma_{3}\cdot \vec Q\over   \vec Q^{2}+m_{\pi}^{2}}
\end{equation}
 and
the other to a pion exchange between particles 1 and 2,
\begin{equation}
\label{30b}
w_{4}^{(1)}|_{3}^{a}=F^{2}(\vec {Q'}^{2})\ 
{\vec \sigma_{2}\cdot \vec {Q'}\over   \vec {Q'}^{2}+m_{\pi}^{2}}
\end{equation}
Consequently one has to evaluate
\begin{eqnarray}
\label{30}
\nonumber
M_{a}^{J,2}&\equiv&
\Bra 2{pq\alpha_{J}}F^{2}(\vec Q^{2})\ 
{\vec \sigma_{3}\cdot \vec Q\over   \vec
  Q^{2}+m_{\pi}^{2}}\ket{p_{1}q_{1}\alpha_{1J}}_{2}\\
M_{a}^{J,3}&\equiv&
\Bra 3{p_{2}q_{2}\alpha_{2J}}F^{2}(\vec {Q'}^{2})\ 
{\vec \sigma_{2}\cdot \vec {Q'}\over   \vec
  {Q'}^{2}+m_{\pi}^{2}}\ket{p'q'\alpha_{J}'}_{3}
\end{eqnarray}

The states $\alpha_{J}$ from eq.~(\ref{27}) couple the total angular momenta
of the two-body subsystem and the third particle:
\begin{equation}
\label{31}
\ket{\alpha_{J}}=\{\ket{\alpha_{j}}\ket{\alpha_{I}}\}^{JM}
\equiv \left\{\ket{(ls)j}\ket{(\lambda {1\over 2})I}\right\}^{JM}
\end{equation}
Further the pion momentum $\vec Q$ from Fig.~\ref{V4(1)} is expressed in terms
of the Jacobi momenta \cite{Buch} in the $(13)$ subsystem as 
\begin{equation}
\label{32}
\vec {Q}=\vec p - \vec p_{1}
\end{equation}
Apparently $M_{a}^{J,2}$ is diagonal in the quantum numbers of the spectator
nucleon, and we find
\begin{eqnarray}
\label{33}
\nonumber
M_{a}^{J,2}&=&{\delta(q-q_{1})\over q^{2}}\ \delta_{\lambda\lambda_{1}}\
\delta_{II_{1}}\ \sum_{mm_{1}} C(jm\ IM-m,JM)\ 
C(j_{1}m_{1}\ I_{1}M_{1}-m_{1},J_{1}M_{1})\\
&\times& {\Bra 2{pjm}
{\vec \sigma_{3}\cdot (\vec p-\vec p_{1})\over 
(\vec p-\vec p_{1})^{2}+m_{\pi}^{2}}\
F^{2}((\vec p-\vec p_{1})^{2})\ket{p_{1}j_{1}m_{1}}_{2}
}
\end{eqnarray}

The scalar $\vec \sigma_{3}\cdot (\vec p-\vec p_{1})$ formed out of two tensor
operators of rank 1 can be decomposed using standard techniques \cite{Buch} as
\begin{eqnarray}
\label{35}
\nonumber
\vec \sigma_{3}\cdot (\vec p - \vec p_{1}) &=& -\sqrt 3\ 
\{ \sigma (3),(\vec p - \vec p_{1})_{1}\}^{00}\\
&=& -4\pi\ \sum_{a+b=1} p^{a}\ (-p_{1})^{b}\
\{ \sigma (3),\Y_{ab}^{1}(\hat p,\hat p_{1})\}^{00}
\end{eqnarray}
As in eq.~(\ref{12a}) the curly brackets denote angular momentum coupling and
$\Y$ the product of two spherical harmonics coupled to angular momentum
1. Note that in contrast to eq.~(\ref{12}) there arises no inverse power of
$|\vec p-\vec p_{1}|$, which would lead to an artificial singularity with the
fatal numerical consequences described in section \ref{old}.

Next we expand the pion propagator and form factors into Legendre polynomials
depending on $x\equiv \hat p \cdot \hat p_{1}$. We rewrite
\begin{equation}
\label{37}
f(x)\equiv{1\over (\vec p - \vec p_{1})^{2}+m_{\pi}^{2}}\
\left( {\Lambda^{2}-m_{\pi}^{2}\over 
\Lambda^{2}+(\vec p - \vec p_{1})^{2}}\right)^{2} 
\end{equation}
as
\begin{equation}
\label{37a}
f(x)= {(\Lambda^{2}-m_{\pi}^{2})^{2}\over (2pp_{1})^{3}}\
\left[ {1\over (B_{\Lambda}-B_{m_{\pi}})^{2}}\
\left( {1\over B_{m_{\pi}}-x}-{1\over B_{\Lambda}-x}\right)
-{1\over B_{\Lambda}-B_{m_{\pi}}}\ {1\over (B_{\Lambda}-x)^{2}}
\right] 
\end{equation}
with
\begin{eqnarray}
\label{38}
\nonumber
B_{\Lambda}&\equiv& {p^{2}+p_{1}^{2}+\Lambda^{2}\over 2pp_{1}}\\
B_{m_{\pi}}&\equiv& {p^{2}+p_{1}^{2}+m_{\pi}^{2}\over 2pp_{1}}
\end{eqnarray}
It results
\begin{equation}
\label{39}
f(x)=\sum_{\overline l=0}^{\infty} {\hat {\overline l}\over 2}\ P_{\overline l}(x)\ H_{\overline l}(p,p_{1})
\end{equation}
with
\begin{equation}
\label{40}
H_{\overline l}(p,p_{1})={1\over pp_{1}}\ (Q_{\overline l}(B_{m_{\pi}})-Q_{\overline l}(B_{\Lambda}))\
+{\Lambda^{2}-m_{\pi}^{2}\over 2(pp_{1})^{2}}\ Q'_{\overline l}(B_{\Lambda})
\end{equation}

These steps lead to the intermediate result for the matrix element in
eq.~(\ref{33}) 
\begin{eqnarray}
\label{41}
\nonumber
M_{a}^{j,2}&\equiv&
{\Bra 2{pjm}
{\vec \sigma_{3}\cdot (\vec p-\vec p_{1})\over 
(\vec p-\vec p_{1})^{2}+m_{\pi}^{2}}\
F^{2}((\vec p-\vec p_{1})^{2})\ket{p_{1}j_{1}m_{1}}_{2}
}\\
&=&-{(4\pi )^{2}\over 2}\
\Bra 2{pjm}\sum_{a+b=1} p^{a}\ (-p_{1})^{b}\
\{ \sigma (3),\Y_{ab}^{1}(\hat p,\hat p_{1})\}^{00}\\
&\times& \sum_{\overline l=0}^{\infty} (-)^{\overline l}\ \sqrt{\hat {\overline l}}\
H_{\overline l}(p,p_{1})\ \Y_{\overline l\overline l}^{00}(\hat p,\hat p_{1}) 
\ket{p_{1}j_{1}m_{1}}_{2}
\end{eqnarray}
or
\begin{eqnarray}
\label{42}
\nonumber
M_{a}^{j,2}&=&-2\pi \sqrt 3 \sum_{a+b=1} p^{a}\ p_{1}^{b}\
\sum_{\overline l=0}^{\infty} \hat {\overline l}\ H_{\overline l}(p,p_{1})\
\sum_{i_{1}i_{2}} (-)^{i_{2}}\\ 
\nonumber
&\times& \left\{ \matrix{a&b&1\cr i_{2}&i_{1}&\overline l}\right\} \
C(a0\ \overline l0,i_{1}0)\ C(b0\ \overline l0,i_{2}0)\\
&\times& \Bra 2{jm}
\{ \sigma (3),\Y_{i_{1}i_{2}}^{1}(\hat p,\hat p_{1})\}^{00}\
\ket{j_{1}m_{1}}_{2}
\end{eqnarray}

This can be expressed in
terms of reduced matrix elements:
\begin{eqnarray}
\label{43}
\nonumber
\Bra 2{jm}
\{ \sigma (3),\Y_{i_{1}i_{2}}^{1}(\hat p,\hat p_{1})\}^{00}\
\ket{j_{1}m_{1}}_{2}
&=&\delta_{jj_{1}}\ \delta_{mm_{1}}\ {\sqrt{\hat l\hat s}\over \sqrt 3}\
(-1)^{1+j+l+s_{1}}\
\left\{ \matrix{l_{1}&s_{1}&j\cr s&l&1}\right\} \\
&\times&
\Braa 2{l} \Y_{i_{1}i_{2}}^{1}(\hat p,\hat p_{1})\kett{l_{1}}_{2}\
\Braa 2s \sigma (3) \kett{s_{1}}_{2}
\end{eqnarray}
with
\begin{equation}
\label{45}
\Braa 2s \sigma (3) \kett{s_{1}}_{2}
= \sqrt 6\ \sqrt{\hat {s_{1}}}\ (-)^{s+1}\
\left\{ \matrix{{1\over 2}&{1\over 2}&s_{1}\cr 1&s&{1\over 2}}\right\}
\end{equation}
and
\begin{equation}
\label{46}
\Braa 2{l} \Y_{i_{1}i_{2}}^{L}(\hat p,\hat p_{1})\kett{l_{1}}_{2}
=\delta_{i_{1}l}\ \delta_{i_{2}l_{1}}\ (-)^{L-l}\
\sqrt{\hat L\over \hat l}
\end{equation}
In view of later use the expression (\ref{46}) is formulated for general $L$. 
Note we define the reduced matrix element as
\begin{equation}
\label{44}
C(Lm-m_{1}\ l_{1}m_{1},lm)\
\braa l \O^{L} \kett{l_{1}}
\equiv \bra{lm} \O^{LM} \ket{l_{1}m_{1}}
\end{equation}
The result for $M_{a}^{j,2}$ is
\begin{eqnarray}
\label{47}
\nonumber
M_{a}^{j,2}&=&6\pi\ \sqrt 2\ \delta_{jj_{1}}\ \delta_{mm_{1}}\
(-)^{j+s+s_{1}+l_{1}}\ \sqrt{\hat s\hat s_{1}}\
\left\{ \matrix{{1\over 2}&{1\over 2}&s_{1}\cr 1&s&{1\over 2}}\right\} \
\left\{ \matrix{l_{1}&s_{1}&j\cr s&l&1}\right\} \\
&\times&\sum_{a+b=1} p^{a}\ p_{1}^{b}\
\sum_{\overline l=0}^{\infty} \hat {\overline l}\ H_{\overline l}(p,p_{1})\
\left\{ \matrix{a&b&1\cr l_{1}&l&\overline l}\right\} \
C(a0\ \overline l0,l0)\ C(b0\ \overline l0,l_{1}0)
\end{eqnarray}

Finally we put this expression into eq.~(\ref{33}) for $M_{a}^{J,2}$. This
allows the summation over the magnetic quantum numbers $m$ and $m_{1}$.
We also can perform the sums over $a+b=1$ and $\overline l$.
We end up with
\begin{eqnarray}
\label{48}
\nonumber
M_{a}^{J,2}&=&{\delta(q-q_{1})\over q^{2}}\ \delta_{jj_{1}}\ 
\delta_{\lambda\lambda_{1}}\ \delta_{II_{1}}\
\delta_{JJ_{1}}\ \delta_{MM_{1}}\ \delta_{|l-l_{1}|,1}\\
\nonumber
&\times& 2\pi\ \sqrt 6\ (-)^{j+1+s+s_{1}}\ \sqrt{\hat s\hat s_{1}}\
\left\{ \matrix{{1\over 2}&{1\over 2}&s_{1}\cr 1&s&{1\over 2}}\right\} \
\left\{ \matrix{l_{1}&s_{1}&j\cr s&l&1}\right\} \\
&\times& \sqrt{\max{(l,l_{1})}}\ 
(p\ H_{l_{1}}(p,p_{1})-p_{1}\ H_{l}(p,p_{1}))\
(-)^{\max{(l,l_{1})}}
\end{eqnarray}
This result for $M_{a}^{J,2}$ is a very simple expression, in contrast to
the $a$-term using the old PWD (see \cite{CG81} and
\cite{PhD}). The complicated structure of the 3NF
in our new PWD is hidden in the three permutation operators of
eqs.~(\ref{29e}) and (\ref{29a}). The latter ones are, however, well under
control algebraically (see 
appendix \ref{perm}) and numerically. 
The $M_{a}^{J,3}$ can be gained out of $M_{a}^{J,2}$ by carefully replacing
quantities with the result
\begin{equation}
\label{52}
M_{a}^{J,2}(p\alpha,p_{1}\alpha_{1})=(-)^{1+s+s_{1}}\
M_{a}^{J,3}(p_{1}\alpha_{1},p\alpha )
\end{equation}

The splitting of the operator $v_{4}^{(1)}$ introduces parity changes for the
individual terms $M_{a}^{J,2}$ and $M_{a}^{J,3}$, as is obvious from
eq.~(\ref{48}). Their product, however, conserves parity, as it should.

The c-term is very similar to the a-term, only $\vec Q^{2}$ and ${\vec
  Q}'^{2}$ factors occur additionally:
\begin{eqnarray}
\label{54}
\nonumber
M_{c}^{J,2}&\equiv&
\Bra 2{pq\alpha_{J}}\vec Q^{2}\ F^{2}(\vec Q^{2})\ 
{\vec \sigma_{3}\cdot \vec Q\over   \vec
  Q^{2}+m_{\pi}^{2}}\ket{p_{1}q_{1}\alpha_{1J}}_{2}\\
M_{c}^{J,3}&\equiv&
\Bra 3{p_{2}q_{2}\alpha_{2J}}\vec {Q'}^{2}\ F^{2}(\vec {Q'}^{2})\ 
{\vec \sigma_{2}\cdot \vec {Q'}\over   \vec
  {Q'}^{2}+m_{\pi}^{2}}\ket{p'q'\alpha_{J}'}_{3}
\end{eqnarray}
This leads to a replacement of $H$ by
\begin{equation}
\label{56}
\tilde H_{\overline l}(p,p_{1})=
-{m_{\pi}^{2}\over pp_{1}}\
(Q_{\overline l}(B_{m_{\pi}})-Q_{\overline l}(B_{\Lambda}))
-{\Lambda^{2}-m_{\pi}^{2}\over 2(pp_{1})^{2}}\
\Lambda^{2}\ Q'_{\overline l}(B_{\Lambda})
\end{equation}
in eq.~(\ref{48}) and provides $M_{c}^{J,2}$ and $M_{c}^{J,3}$. The
symmetry relation eq.~(\ref{52}) is also valid for $M_{c}^{J,2}$ and
$M_{c}^{J,3}$. 

The b-term in eq.~(\ref{24a}) has the scalar product $\vec Q\cdot \vec {Q'}$ on
top of the structure of the a-term.
That scalar product $\vec Q\cdot \vec {Q'}$ can be rewritten as:
\begin{eqnarray}
\label{60}
\nonumber
\vec Q\cdot \vec {Q'}&=&
-\sqrt 3\ \{ Q_{1},Q'_{1}\}^{00}
=-\sqrt 3\ {4\pi\over 3}\ QQ'\ \Y_{11}^{00}(\hat Q,\hat {Q'})\\
&=&{4\pi\over 3}\ QQ'\ \sum_{\mu} (-)^{\mu}\
Y_{1}^{\mu}(\hat Q)\ Y_{1}^{-\!\!\mu}(\hat {Q'})
\end{eqnarray}
This leads to the amplitudes
\begin{eqnarray}
\label{62}
\nonumber
M_{b}^{J,2}&\equiv& \sqrt{4\pi\over 3}\
\Bra 2{pq\alpha_{J}}Q\ Y_{1}^{\mu}(\hat {Q})
F^{2}(\vec Q^{2})\ 
{\vec \sigma_{3}\cdot \vec Q\over   \vec
  Q^{2}+m_{\pi}^{2}}\ket{p_{1}q_{1}\alpha_{1J}}_{2}\\
M_{b}^{J,3}&\equiv& \sqrt{4\pi\over 3}\
\Bra 3{p_{2}q_{2}\alpha_{2J}}Q'\ Y_{1}^{-\!\!\mu}(\hat {Q'})
F^{2}(\vec {Q'}^{2})\ 
{\vec \sigma_{2}\cdot \vec {Q'}\over   \vec
  {Q'}^{2}+m_{\pi}^{2}}\ket{p'q'\alpha_{J}'}_{3}
\end{eqnarray}

Now similarly to eq.(\ref{33}) one has
\begin{eqnarray}
\label{62a}
\nonumber
M_{b}^{J,2}&=&{\delta(q-q_{1})\over q^{2}}\ \delta_{\lambda\lambda_{1}}\
\delta_{II_{1}}\ \sum_{mm_{1}} C(jm\ IM-m,JM)\ 
C(j_{1}m_{1}\ I_{1}M_{1}-m_{1},J_{1}M_{1})\\
&\times& {\sqrt{4\pi\over 3}\ \Bra 2{pjm}
{\vec \sigma_{3}\cdot (\vec p-\vec p_{1})\over 
(\vec p-\vec p_{1})^{2}+m_{\pi}^{2}}\
F^{2}((\vec p-\vec p_{1})^{2})\ 
|\vec p-\vec p_{1}|\ Y_{1}^{\mu}(\widehat {\vec p-\vec p_{1}})
\ket{p_{1}j_{1}m_{1}}_{2}
}
\end{eqnarray}

Its evaluation requires the following recoupling:
\begin{eqnarray}
\label{63}
\nonumber
&&Y_{1}^{\mu}(\widehat{\vec p-\vec p_{1}})\
\vec \sigma (3)\cdot (\vec p-\vec p_{1})
= -\sqrt{4\pi}\ |\vec p-\vec p_{1}|\
\left\{ Y_{1}(\widehat{\vec p-\vec p_{1}}),
\{ \sigma (3),Y_{1}(\widehat{\vec p-\vec p_{1}})\}^{0}
\right\}^{1,\mu}\\
\nonumber
&=&{\sqrt{4\pi}\over 3}\ |\vec p-\vec p_{1}|\
\sum_{i_{1}} (-)^{i_{1}+1}\ \sqrt{\hat i_{1}}\
\left\{ \sigma (3),
\Y_{11}^{i_{1}}(\widehat{\vec p-\vec p_{1}},\widehat{\vec p-\vec p_{1}})
\right\}^{1,\mu}\\
\nonumber
&=& |\vec p-\vec p_{1}|\ \sum_{i_{1}} (-)^{i_{1}+1}\ 
C(10\ 10,i_{1}0)\ 
\left\{ \sigma (3), Y_{i_{1}}(\widehat{\vec p-\vec p_{1}})
\right\}^{1,\mu}\\
&=& |\vec p-\vec p_{1}|\ 
\left( {1\over \sqrt 3}\
\left\{ \sigma (3), Y_{0}(\widehat{\vec p-\vec p_{1}})
\right\}^{1,\mu}
-\sqrt{2\over 3}\
\left\{ \sigma (3), Y_{2}(\widehat{\vec p-\vec p_{1}})
\right\}^{1,\mu}
\right)
\end{eqnarray}
As an intermediate step using eq.~(\ref{11}) we get
\begin{eqnarray}
\label{64}
\nonumber
M_{b}^{j,2}&\equiv&
{\sqrt{4\pi\over 3}\ \Bra 2{pjm}
{\vec \sigma_{3}\cdot (\vec p-\vec p_{1})\over 
(\vec p-\vec p_{1})^{2}+m_{\pi}^{2}}\
F^{2}(\vec p-\vec p_{1})^{2}\
|\vec p-\vec p_{1}|\ Y_{1}^{\mu}(\widehat {\vec p-\vec p_{1}})
\ket{p_{1}j_{1}m_{1}}_{2}
}\\
\nonumber
&=&{4\pi\over 3}\
\Bra 2{pjm}(\vec p-\vec p_{1})^{2}\ 
{F^{2}(\vec p-\vec p_{1})^{2}\over (\vec p-\vec p_{1})^{2}+m_{\pi}^{2}}\
\left\{ \sigma (3), \Y_{00}^{0}(\widehat{\vec p-\vec p_{1}})
\right\}^{1,\mu}\ket{p_{1}j_{1}m_{1}}_{2}\\
\nonumber
&-&{4\pi\over 3}\ 4\sqrt {15}\
\Bra 2{pjm}
{F^{2}(\vec p-\vec p_{1})^{2}\over (\vec p-\vec p_{1})^{2}+m_{\pi}^{2}}\\
&\times&
\sum_{a+b=2} {p^{a}(-p_{1})^{b}\over \sqrt{\hat a!\hat b!}}\
\left\{ \sigma (3), \Y_{ab}^{2}(\widehat{\vec p-\vec p_{1}})
\right\}^{1,\mu}\ket{p_{1}j_{1}m_{1}}_{2}
\end{eqnarray}
The expansion into Legendre polynomials for the first term is the same as for
the c-term and for the second term as for the a-term. We get
\begin{eqnarray}
\label{65}
\nonumber
M_{b}^{j,2}&=&{8\pi^{2}\over 3}\
\sum_{\overline l} (-)^{\overline l}\ \sqrt{\hat {\overline l}}\ \tilde H_{\overline l}(p,p_{1})\
\Bra 2{jm}\Y_{\overline l\overline l}^{00}(\hat p,\hat p_{1})\
\left\{ \sigma (3), \Y_{00}^{0}(\hat p,\hat p_{1})
\right\}^{1,\mu}\ket{j_{1}m_{1}}_{2}\\
\nonumber
&-& {8\pi^{2}\over 3}\ 4\sqrt{15}\
\sum_{\overline l} (-)^{\overline l}\ \sqrt{\hat {\overline l}}\ H_{\overline l}(p,p_{1})\\
&\times & 
\sum_{a+b=2} {p^{a}(-p_{1})^{b}\over \sqrt{\hat a!\hat b!}}\
\Bra 2{jm}\Y_{\overline l\overline l}^{00}(\hat p,\hat p_{1})\
\left\{ \sigma (3), \Y_{ab}^{2}(\hat p,\hat p_{1})
\right\}^{1,\mu}\ket{j_{1}m_{1}}_{2}
\end{eqnarray}

To proceed further we have to recouple
\begin{eqnarray}
\label{66}
\nonumber
&&\Y_{\overline l\overline l}^{00}(\hat p,\hat p_{1})\
\left\{ \sigma (3), \Y_{ab}^{2}(\hat p,\hat p_{1})
\right\}^{1,\mu}\\
\nonumber
&=& {1\over 4\pi}\ \sqrt{\hat a\hat b\hat {\overline l}} 
\sum_{i_{3}i_{4}} (-)^{i_{4}+a+\overline l}\ 
\left\{ \matrix{b&a&2\cr i_{3}&i_{4}&\overline l}\right\} \\
&\times& 
C(a0\ \overline l0,i_{3}0)\ C(b0\ \overline l0,i_{4}0)\
\left\{ \sigma (3), \Y_{i_{3}i_{4}}^{2}(\hat p,\hat p_{1})
\right\}^{1,\mu}
\end{eqnarray}
and
\begin{equation}
\label{67}
\nonumber
\Y_{\overline l\overline l}^{00}(\hat p,\hat p_{1})\
\left\{ \sigma (3), \Y_{00}^{0}(\hat p,\hat p_{1})
\right\}^{1,\mu}
={1\over 4\pi}\ 
\left\{ \sigma (3), \Y_{\overline l\overline l}^{0}(\hat p,\hat p_{1})
\right\}^{1,\mu}
\end{equation}
which leads
directly to the following two matrix elements:
\begin{eqnarray}
\label{68}
\nonumber
&&\Bra 2{jm}\left\{ \sigma (3), \Y_{i_{3}i_{4}}^{2}(\hat p,\hat p_{1})
\right\}^{1,\mu} \ket{j_{1}m_{1}}_{2}\\
&=&\delta_{i_{3}l}\ \delta_{i_{4}l_{1}}\
C(1\mu\ j_{1}m_{1},jm)\ 3\sqrt{10}\
(-)^{s-l+1}\ \sqrt{\hat j_{1}\hat s\hat s_{1}}\
\left\{ \matrix{{1\over 2}&{1\over 2}&s_{1}\cr
                1&s&{1\over 2}}\right\} \
\left\{ \matrix{2&1&1\cr l_{1}&s_{1}&j_{1}\cr l&s&j}\right\}
\end{eqnarray}
and 
\begin{eqnarray}
\label{69}
\nonumber
&&\Bra 2{jm}\left\{ \sigma (3), \Y_{\overline l\overline l}^{0}(\hat p,\hat p_{1})
\right\}^{1,\mu} \ket{j_{1}m_{1}}_{2}\\
&=&\delta_{\overline ll}\ \delta_{\overline ll_{1}}\
C(1\mu\ j_{1}m_{1},jm)\ \sqrt 6\ (-)^{j_{1}}\
{\sqrt{\hat j_{1}\hat s\hat s_{1}}\over \sqrt{\hat l_{1}}}\
\left\{ \matrix{{1\over 2}&{1\over 2}&s_{1}\cr
                1&s&{1\over 2}}\right\} \
\left\{ \matrix{j&j_{1}&1\cr s_{1}&s&l}\right\}
\end{eqnarray}
and then to
\begin{eqnarray}
\label{70}
\nonumber
M_{b}^{j,2}&=& C(1\mu\ j_{1}m_{1},jm)\\
\nonumber
&\times& \biggl[
\delta_{ll_{1}}\ {2\pi\over 3}\ \sqrt 6\
(-)^{l+j_{1}}\ \sqrt{\hat j_{1}\hat s\hat s_{1}}\ \tilde H_{l}(p,p_{1})\
\left\{ \matrix{{1\over 2}&{1\over 2}&s_{1}\cr
                1&s&{1\over 2}}\right\} \
\left\{ \matrix{j&j_{1}&1\cr s_{1}&s&l}\right\} \\
\nonumber
&-&40\pi\ \sqrt 6\ (-)^{s+1}\ 
\sqrt{\hat j_{1}\hat s\hat s_{1}}\ 
\left\{ \matrix{{1\over 2}&{1\over 2}&s_{1}\cr
                1&s&{1\over 2}}\right\} \
\left\{ \matrix{2&1&1\cr l_{1}&s_{1}&j_{1}\cr l&s&j}\right\}\\
&\times&
\sum_{\overline l} \hat {\overline l}\ H_{\overline l}(p,p_{1})\
\sum_{a+b=2} {p^{a}\ p_{1}^{b}\over \sqrt{(2a)!\ (2b)!}}\
\left\{ \matrix{b&a&2\cr l&l_{1}&\overline l}\right\} \
C(a0\ \overline l0,l0)\ C(b0\ \overline l0,l_{1}0) \biggr]
\end{eqnarray}
The summation over the magnetic quantum numbers in
eq.~(\ref{62a}) using the $m$, $m_{1}$ dependence from eq.~(\ref{70}) can be
done analytically:  
\begin{eqnarray}
\label{71}
\nonumber
&&\sum_{mm_{1}} C(jm\ IM-m,JM)\ 
C(j_{1}m_{1}\ I_{1}M_{1}-m_{1},J_{1}M_{1})\ 
C(1\mu\ j_{1}m_{1},jm)\\
&=&(-)^{1+j_{1}+I+J}\ \sqrt{\hat j\hat J_{1}}\
C(1\mu\ J_{1}M_{1},JM)\
\left\{ \matrix{1&j_{1}&j\cr I&J&J_{1}}\right\}
\end{eqnarray}
and we get for $M_{b}^{J,2}$
\begin{eqnarray}
\label{72}
\nonumber
M_{b}^{J,2}&=&{\delta(q-q_{1})\over q^{2}}\ \delta_{\lambda\lambda_{1}}\
\delta_{II_{1}}\ C(1\mu\ J_{1}M_{1},JM)\\
\nonumber
&\times& (-)^{I+J}\ \sqrt{\hat j\hat j_{1}\hat s\hat s_{1}\hat J_{1}}\
\left\{ \matrix{{1\over 2}&{1\over 2}&s_{1}\cr
                1&s&{1\over 2}}\right\} \
\left\{ \matrix{1&j_{1}&j\cr I&J&J_{1}}\right\} \\
\nonumber
&\times&\biggl[
\delta_{ll_{1}}\ {2\pi\over 3}\ \sqrt 6\
(-)^{l+1}\ \tilde H_{l}(p,p_{1})\
\left\{ \matrix{j&j_{1}&1\cr s_{1}&s&l}\right\} \\
\nonumber
&-&40\pi\ \sqrt 6\ (-)^{s+j_{1}}\ 
\left\{ \matrix{2&1&1\cr l_{1}&s_{1}&j_{1}\cr l&s&j}\right\}\\
&\times&
\sum_{\overline l} \hat {\overline l}\ H_{\overline l}(p,p_{1})\
\sum_{a+b=2} {p^{a}\ p_{1}^{b}\over \sqrt{(2a)!\ (2b)!}}\
\left\{ \matrix{b&a&2\cr l&l_{1}&\overline l}\right\} \
C(a0\ \overline l0,l0)\ C(b0\ \overline l0,l_{1}0) \biggr]
\end{eqnarray}

The evaluation of $M_{b}^{J,3}$ proceeds along similar lines or can be
abbreviated by judicious replacements and using symmetries of 6j- and
9j-symbols. The result is
\begin{eqnarray}
\label{73}
\nonumber
M_{b}^{J,3}&=&{\delta(q'-q_{2})\over q'^{2}}\ \delta_{\lambda'\lambda_{2}}\
\delta_{I'I_{2}}\ C(1-\!\!\mu\ J'M',J_{2}M_{2})\\
\nonumber
&\times& (-)^{I'+J_{2}+s'-s_{2}}\ 
\sqrt{\hat j'\hat j_{2}\hat s'\hat s_{2}\hat J'}\
\left\{ \matrix{{1\over 2}&{1\over 2}&s_{2}\cr
                1&s'&{1\over 2}}\right\} \
\left\{ \matrix{1&j_{2}&j'\cr I'&J'&J_{2}}\right\} \\
\nonumber
&\times&\biggl[
\delta_{l'l_{2}}\ {2\pi\over 3}\ \sqrt 6\
(-)^{l'+1}\ \tilde H_{l'}(p'p_{2})\
\left\{ \matrix{j'&j_{2}&1\cr s_{2}&s'&l'}\right\} \\
\nonumber
&-&40\pi\ \sqrt 6\ (-)^{s'+j_{2}}\ 
\left\{ \matrix{2&1&1\cr l_{2}&s_{2}&j_{2}\cr l'&s'&j'}\right\}\\
&\times&
\sum_{\overline l} \hat {\overline l}\ H_{\overline l}(p'p_{2})\
\sum_{a+b=2} {p'^{a}\ p_{2}^{b}\over \sqrt{(2a)!\ (2b)!}}\
\left\{ \matrix{b&a&2\cr l'&l_{2}&\overline l}\right\} \
C(a0\ \overline l0,l'0)\ C(b0\ \overline l0,l_{2}0) \biggr]
\end{eqnarray}

Now, looking back to eq.~(\ref{60}), we can perform the sum over $\mu$:
\begin{equation}
\label{74}
\sum_{\mu} (-)^{\mu}\ C(1\mu\ J_{1}m_{1},JM)\ C(1-\!\!\mu\ J'M',J_{1}M_{1})
=\delta_{JJ'}\ \delta_{MM'}\ (-)^{J_{1}-J}\ \sqrt{\hat J_{1}\over \hat J}
\end{equation}
Thereby we used $J_{1}=J_{2}$ and $M_{1}=M_{2}$, which results from the fact
that the permutation operator standing between $M_{b}^{J,2}$ and $M_{b}^{J,3}$
(see eq.~(\ref{29a}))
conserves $J$ and $M$. Thus despite the fact that $M_{b}^{J,2}$ and
$M_{b}^{J,3}$ do not conserve $J$, as is obvious from the Clebsch-Gordan
coefficients in eq.~(\ref{74}), the total expression $v_{4}^{(1)}|_{b}$
does it, as expected. The parity, however, is conserved by the individual
quantities $M_{b}^{J,2}$ and $M_{b}^{J,3}$.

Having performed the $\mu$-summation reduced $M$-quantities occur, which are
defined as
\begin{eqnarray}
\label{75}
\nonumber
M_{b}^{J,2}&\equiv& 
C(1\mu\ J_{1}M_{1},JM)\ (-)^{J_{1}-J}\ \overline M_{b}^{J,2}\\
M_{b}^{J,3}&\equiv& 
C(1-\!\!\mu\ J'M',J_{2}M_{2})\ \sqrt{\hat J'\over \hat J_{2}}\ 
\overline M_{b}^{J,3}
\end{eqnarray}
and which obey the following symmetry relation
\begin{equation}
\label{76}
\overline M_{b}^{J,2} (p\alpha ,p_{1}\alpha_{1})=
(-)^{s_{1}-s}\ \overline M_{b}^{J,3}(p_{1}\alpha_{1},p\alpha )
\end{equation}

Again we see that the expressions using this new PWD, now for the $b$-term,
are much simpler than the expressions of the old PWD, see
\cite{CG81} and \cite{PhD}. 

The last term, the $d$-term, contains the operator
$\vec \sigma_{1}\cdot \vec
{Q}\times \vec {Q'}$ (see eq.~(\ref{24a})):
\begin{eqnarray}
\label{79}
\nonumber
\vec \sigma_{1}\cdot \vec {Q}\times \vec {Q'}
&=&\vec \sigma_{1}\cdot (i\sqrt 2)\ \left\{ Q_{1},Q'_{1}\right\}^{1}
=i\sqrt 6\ {4\pi\over 3}\ QQ'\
\left\{ \sigma (1), 
\left\{ Y_{1}(\hat Q),Y_{1}(\hat {Q'})\right\}^{1} \right\}^{0,0}\\
\nonumber
&=&-i\sqrt 2\ {4\pi\over 3}\ QQ'
\sum_{\mu} (-)^{\mu}\ 
\left\{ \sigma (1),Y_{1}(\hat Q)\right\}^{1,-\!\!\mu}\ Y_{1}^{\mu}(\hat {Q'})\\
&=&i\sqrt 2\ {4\pi\over 3}\ QQ'
\sum_{\mu} (-)^{\mu}\ 
\left\{ \sigma (1),Y_{1}(\hat {Q'})\right\}^{1,-\!\!\mu}\ Y_{1}^{\mu}(\hat {Q})
\end{eqnarray}
According to
the two possibilities to couple $\sigma (1)$ 
we have two ways to write down the $d$-term:
\begin{eqnarray}
\label{80}
\nonumber
\Bra 1{pq\alpha}v_{4}^{(1)}|^{d}\ket {p'q'\alpha'}_{3}
&=&P_{1\leftrightarrow 2}\ 
\sum_{\mu} (-)^{\mu}\ M_{d}^{J,2}\
P_{2\leftrightarrow 3}\ M_{b}^{J,3}\\
&=&P_{1\leftrightarrow 2}\ 
\sum_{\mu} (-)^{\mu}\ M_{b}^{J,2}\
P_{2\leftrightarrow 3}\ M_{d}^{J,3}
\end{eqnarray}
with
\begin{eqnarray}
\label{81}
\nonumber
M_{d}^{J,2}&\equiv& -i\sqrt 2\ \sqrt{4\pi\over 3}\
\Bra 2{pq\alpha_{J}}Q\ \left\{ \sigma (1),Y_{1}(\hat Q)\right\}^{1\mu}\ 
F^{2}(\vec Q^{2})\ 
{\vec \sigma_{3}\cdot \vec Q\over   \vec
  Q^{2}+m_{\pi}^{2}}\ket{p_{1}q_{1}\alpha_{1J}}_{2}\\
M_{d}^{J,3}&\equiv& i\sqrt 2\ \sqrt{4\pi\over 3}\
\Bra 3{p_{2}q_{2}\alpha_{2J}}Q'\ 
\left\{ \sigma (1),Y_{1}(\hat {Q'})\right\}^{1,-\!\!\mu}\ 
F^{2}(\vec {Q'}^{2})\ 
{\vec \sigma_{2}\cdot \vec {Q'}\over   \vec
  {Q'}^{2}+m_{\pi}^{2}}\ket{p'q'\alpha_{J}'}_{3}
\end{eqnarray}
The permutation operators $P_{1\leftrightarrow 2}$ and $P_{2\leftrightarrow
  3}$ in eq.~(\ref{80}) stand for the recoupling matrix elements (see
eq.~(\ref{29a})) and we dropped the iso-spin factor $I_{4}^{(1)}$.
We see that $M_{d}^{J,2}$ and $M_{d}^{J,3}$ are purely imaginary. Together
with the imaginary iso-spin factor of eq.~(\ref{29}), the matrix element is of
course real.

For $M_{d}^{J,2}$ we have to recouple the operator
\begin{eqnarray}
\label{82}
\nonumber
&&\vec \sigma_{3}\cdot \vec Q\ \left\{ \sigma (1),Y_{1}(\hat Q)\right\}^{1\mu}
= \vec \sigma_{3}\cdot (\vec p-\vec p_{1})\
\left\{ \sigma (1),Y_{1}(\widehat {\vec p-\vec p_{1}})\right\}^{1\mu}\\
\nonumber
&=&-\sqrt 3\ \sqrt{4\pi\over 3}\ |\vec p-\vec p_{1}|\
\left\{ 
\left\{ \sigma (3),Y_{1}(\widehat {\vec p-\vec p_{1}})\right\}^{0},
\left\{ \sigma (1),Y_{1}(\widehat {\vec p-\vec p_{1}})\right\}^{1}
\right\}^{1\mu}\\
\nonumber
&=&\sqrt 3\ |\vec p-\vec p_{1}|\
\sum_{i_{1}i_{2}} (-)^{i_{2}}\ \sqrt{\hat i_{1}}\ C(10\ 10,i_{2}0)\
\left\{ \matrix{i_{2}&i_{1}&1\cr 1&1&1}\right\}\\
\nonumber
&\times& \left\{ 
\left\{ \sigma (1),\sigma (3)\right\}^{i_{1}},
Y_{i_{2}}(\widehat {\vec p-\vec p_{1}})
\right\}^{1\mu}\\
\nonumber
&=&{1\over \sqrt 3}\ |\vec p-\vec p_{1}|\
\left\{ 
\left\{ \sigma (1),\sigma (3)\right\}^{1},
Y_{0}(\widehat {\vec p-\vec p_{1}})
\right\}^{1\mu}\\
&+& \sqrt 2\ |\vec p-\vec p_{1}|\
\sum_{i_{1}=1,2} \sqrt{\hat i_{1}}\ 
\left\{ \matrix{2&i_{1}&1\cr 1&1&1}\right\}\
\left\{ 
\left\{ \sigma (1),\sigma (3)\right\}^{i_{1}},
Y_{2}(\widehat {\vec p-\vec p_{1}})
\right\}^{1\mu}
\end{eqnarray}
In the last step we inserted explicitly the sum over the two
$i_{2}$-values. Finally we decompose the spherical harmonics and get
\begin{eqnarray}
\label{83a}
\nonumber
&&\vec \sigma_{3}\cdot \vec Q\ 
\left\{ \sigma (1),Y_{1}(\hat Q)\right\}^{1\mu}\\
\nonumber
&=&\sqrt{4\pi\over 3}\ |\vec p-\vec p_{1}|\
\left\{ 
\left\{ \sigma (1),\sigma (3)\right\}^{1},
\Y_{00}^{0}(\hat p,\hat p_{1})
\right\}^{1\mu}\\
\nonumber
&+& 8\sqrt {15\pi}\ {1\over |\vec p-\vec p_{1}|}\
\sum_{i_{1}=1,2} \sqrt{\hat i_{1}}\ 
\left\{ \matrix{2&i_{1}&1\cr 1&1&1}\right\}\\
&\times&
\sum_{a+b=2} {p^{a}\ (-p_{1})^{b}\over \sqrt{\hat a!\hat b!}}\
\left\{ 
\left\{ \sigma (1),\sigma (3)\right\}^{i_{1}},
\Y^{2}_{ab}(\hat p,\hat p_{1})
\right\}^{1\mu}
\end{eqnarray}
which inserted into $M_{d}^{J,2}$ yields:
\begin{eqnarray}
\label{84}
\nonumber
M_{d}^{J,2}&=&
-i\sqrt 2\ {4\pi\over 3}\
\Bra 2{pq\alpha_{J}}
{F^{2}(|\vec p-\vec p_{1}|^{2})\over |\vec p-\vec p_{1}|^{2}+m_{\pi}^{2}}\
|\vec p-\vec p_{1}|^{2}\
\left\{ \left\{ \sigma (1),\sigma (3)\right\}^{1},
\Y_{00}^{0}(\hat p,\hat p_{1})\right\}^{1\mu}
\ket{p_{1}q_{1}\alpha_{1J}}_{2}\\
\nonumber
&-&i16\pi\ \sqrt{10}\
\sum_{i_{1}=1,2} \sqrt{\hat i_{1}}\ 
\left\{ \matrix{2&i_{1}&1\cr 1&1&1}\right\}
\sum_{a+b=2} {p^{a}\ (-p_{1})^{b}\over \sqrt{\hat a!\hat b!}}\\
&\times&
\Bra 2{pq\alpha_{J}}
{F^{2}(|\vec p-\vec p_{1}|^{2})\over |\vec p-\vec p_{1}|^{2}+m_{\pi}^{2}}\
\left\{ 
\left\{ \sigma (1),\sigma (3)\right\}^{i_{1}},
\Y^{2}_{ab}(\hat p,\hat p_{1})
\right\}^{1\mu}
\ket{p_{1}q_{1}\alpha_{1J}}_{2}
\end{eqnarray}
This expression contains no singularity.

It remains to expand
the $x$-dependence in terms of Legendre polynomials, which leads to
\begin{eqnarray}
\label{85}
\nonumber
M_{d}^{J,2}&=&
-i\sqrt 2\ {8\pi^{2}\over 3}\
\sum_{\overline l} \sqrt{\hat {\overline l}}\ (-)^{\overline l}\ 
\tilde H_{\overline l}(p,p_{1})\\
\nonumber
&\times&
\Bra 2{pq\alpha_{J}}
\Y_{\overline l\overline l}^{0}(\hat p,\hat p_{1})\
\left\{ \left\{ \sigma (1),\sigma (3)\right\}^{1},
\Y_{00}^{0}(\hat p,\hat p_{1})\right\}^{1\mu}
\ket{p_{1}q_{1}\alpha_{1J}}_{2}\\
\nonumber
&-&i32\pi^{2}\ \sqrt{10}\
\sum_{i_{1}=1,2} \sqrt{\hat i_{1}}\ 
\left\{ \matrix{2&i_{1}&1\cr 1&1&1}\right\}
\sum_{a+b=2} {p^{a}\ (-p_{1})^{b}\over \sqrt{\hat a!\hat b!}}\
\sum_{\overline l} \sqrt{\hat {\overline l}}\ (-)^{\overline l}\ 
H_{\overline l}(p,p_{1})\\
&\times&
\Bra 2{pq\alpha_{J}}
\Y_{\overline l\overline l}^{0}(\hat p,\hat p_{1})\
\left\{ 
\left\{ \sigma (1),\sigma (3)\right\}^{i_{1}},
\Y^{2}_{ab}(\hat p,\hat p_{1})
\right\}^{1\mu}
\ket{p_{1}q_{1}\alpha_{1J}}_{2}
\end{eqnarray}
and to combine the spherical harmonics
\begin{eqnarray}
\label{86}
\nonumber
&&\Y_{\overline l\overline l}^{00}(\hat p,\hat p_{1})\
\left\{ 
\left\{ \sigma (1),\sigma (3)\right\}^{i_{1}},
\Y^{2}_{ab}(\hat p,\hat p_{1})
\right\}^{1\mu}\\
\nonumber
&=&\left\{ \left\{ 
\left\{ \sigma (1),\sigma (3)\right\}^{i_{1}},
\Y^{2}_{ab}(\hat p,\hat p_{1})
\right\}^{1}
\Y_{\overline l\overline l}^{0}(\hat p,\hat p_{1})\
\right\}^{1\mu}\\
\nonumber
&=&\sum_{i_{3}i_{4}} {1\over 4\pi}\ (-)^{\overline l+i_{4}+a}\
\sqrt{\hat a\hat b\hat {\overline l}}\
C(a0\ \overline l0,i_{3}0)\ C(b0\ \overline l0,i_{4}0)\\
&\times&\left\{ \matrix{a&b&2\cr i_{4}&i_{3}&\overline l}\right\}\
\left\{ 
\left\{ \sigma (1),\sigma (3)\right\}^{i_{1}},
\Y^{2}_{i_{3}i_{4}}(\hat p,\hat p_{1})
\right\}^{1\mu}
\end{eqnarray}
and
\begin{equation}
\label{87}
\nonumber
\Y_{\overline l\overline l}^{00}(\hat p,\hat p_{1})\
\left\{ 
\left\{ \sigma (1),\sigma (3)\right\}^{1},
\Y^{0}_{00}(\hat p,\hat p_{1})
\right\}^{1\mu}\
={1\over 4\pi}\
\left\{ 
\left\{ \sigma (1),\sigma (3)\right\}^{1},
\Y^{0}_{\overline l\overline l}(\hat p,\hat p_{1})
\right\}^{1\mu}
\end{equation}

Thus we have to evaluate two matrix elements. The first is
\begin{eqnarray}
\label{88}
\nonumber
&&\Bra 2{pq\alpha}
\left\{ 
\left\{ \sigma (1),\sigma (3)\right\}^{i_{1}},
\Y^{2}_{i_{3}i_{4}}(\hat p,\hat p_{1})
\right\}^{1\mu}
\ket {p_{1}q_{1}\alpha_{1}}_{2}\\
\nonumber 
&=&{\delta (q-q_{1})\over q^{2}}\ 
\delta_{\lambda\lambda_{1}}\ \delta_{II_{1}}\ 
\sum_{mm_{1}} C(jm\ IM-m,JM)\ C(j_{1}m_{1}\ I_{1}M_{1}-m_{1},J_{1}M_{1})\\
\nonumber
&\times&
C(1\mu\ j_{1}m_{1},jm)\ (-)^{i_{1}+1}\ \sqrt{3\hat j_{1}\hat l\hat s}\
\left\{ \matrix{2&i_{1}&1\cr l_{1}&s_{1}&j_{1}\cr l&s&j} \right\} \\
&\times&
\Braa 2l \Y^{2}_{i_{3}i_{4}}(\hat p,\hat p_{1})\kett{l_{1}}_{2}\
\Braa 2s \left\{ \sigma (1),\sigma (3)\right\}^{i_{1}} \kett{s_{1}}_{2}
\end{eqnarray}
Only the reduced spin matrix element is new:
\begin{equation}
\label{89}
\nonumber
\Braa 2s \left\{ \sigma (1),\sigma (3)\right\}^{i_{1}} \kett{s_{1}}_{2}
=6\ \sqrt{\hat i_{1}\hat s_{1}}\
\left\{ \matrix{1&1&i_{1}\cr 
        {1\over 2}&{1\over 2}&s_{1}\cr
        {1\over 2}&{1\over 2}&s}\right \}
\end{equation}
After summation over $m$ and $m_{1}$ we get
\begin{eqnarray}
\label{90}
\nonumber
&&\Bra 2{pq\alpha}
\left\{ 
\left\{ \sigma (1),\sigma (3)\right\}^{i_{1}},
\Y^{2}_{i_{3}i_{4}}(\hat p,\hat p_{1})
\right\}^{1\mu}
\ket {p_{1}q_{1}\alpha_{1}}_{2}\\
\nonumber 
&=&{\delta (q-q_{1})\over q^{2}}\ 
\delta_{\lambda\lambda_{1}}\ \delta_{II_{1}}\ 
\delta_{i_{3}l}\ \delta_{i_{4}l_{1}}\
6\sqrt {15}\ (-)^{j_{1}-l+I+J+i_{1}}\
\sqrt{\hat j\hat j_{1}\hat s\hat s_{1}\hat J_{1}\hat i_{1}}\\
&\times&
C(1\mu\ J_{1}M_{1},JM)\ 
\left\{ \matrix{1&j_{1}&j\cr I&J&J_{1}}\right\}\
\left\{ \matrix{2&i_{1}&1\cr l_{1}&s_{1}&j_{1}\cr l&s&j} \right\} \
\left\{ \matrix{1&1&i_{1}\cr 
        {1\over 2}&{1\over 2}&s_{1}\cr
        {1\over 2}&{1\over 2}&s}\right \}
\end{eqnarray}

The second matrix element is
\begin{eqnarray}
\label{91}
\nonumber
&&\Bra 2{pq\alpha}
\left\{ 
\left\{ \sigma (1),\sigma (3)\right\}^{1},
\Y^{0}_{\overline l\overline l}(\hat p,\hat p_{1})
\right\}^{1\mu}
\ket {p_{1}q_{1}\alpha_{1}}_{2}\\
\nonumber 
&=&{\delta (q-q_{1})\over q^{2}}\ 
\delta_{\lambda\lambda_{1}}\ \delta_{II_{1}}\ 
\delta_{\overline ll}\ \delta_{\overline ll_{1}}\ 
6\sqrt 3\ (-)^{s+I+J}\
{\sqrt{\hat j\hat j_{1}\hat s\hat s_{1}\hat J_{1}}\over\sqrt{\hat l}}\\
&\times&
C(1\mu\ J_{1}M_{1},JM)\ 
\left\{ \matrix{1&j_{1}&j\cr I&J&J_{1}}\right\}\
\left\{ \matrix{l&s&j\cr 1&j_{1}&s_{1}}\right\}\
\left\{ \matrix{1&1&1\cr 
        {1\over 2}&{1\over 2}&s_{1}\cr
        {1\over 2}&{1\over 2}&s}\right \}
\end{eqnarray}

We end up with
\begin{eqnarray}
\label{92}
\nonumber
M_{d}^{J,2}&=&{\delta (q-q_{1})\over q^{2}}\ 
\delta_{\lambda\lambda_{1}}\ \delta_{II_{1}}\ 
(-)^{I+J+1}\ \sqrt{\hat j\hat j_{1}\hat s\hat s_{1}\hat J_{1}}\
C(1\mu\ J_{1}M_{1},JM)\ 
\left\{ \matrix{1&j_{1}&j\cr I&J&J_{1}}\right\}\\
\nonumber
&\times&
\biggl[ \delta_{ll_{1}}\ i\ 4\pi\ \sqrt 6\ (-)^{l+s}\
\left\{ \matrix{l&s&j\cr 1&j_{1}&s_{1}}\right\}\
\left\{ \matrix{1&1&1\cr 
        {1\over 2}&{1\over 2}&s_{1}\cr
        {1\over 2}&{1\over 2}&s}\right \}\
\tilde H_{l}(p,p_{1})\\
\nonumber
&+&i\ 240\pi\ \sqrt6\ (-)^{j_{1}}\
\sum_{i_{1}} (-)^{i_{1}}\ \hat i_{1}\
\left\{ \matrix{2&i_{1}&1\cr 1&1&1}\right\}\
\left\{ \matrix{2&i_{1}&1\cr l_{1}&s_{1}&j_{1}\cr l&s&j} \right\} \
\left\{ \matrix{1&1&i_{1}\cr 
        {1\over 2}&{1\over 2}&s_{1}\cr
        {1\over 2}&{1\over 2}&s}\right \}\\
&\times&
\sum_{a+b=2} {p^{a}\ p_{1}^{b}\over \sqrt{(2a)!\ (2b)!}}\
\sum_{\overline l} \hat {\overline l}\ H_{\overline l}(p,p_{1})\
\left\{ \matrix{a&b&2\cr l_{1}&l&\overline l}\right\} \
C(a0\ \overline l0,l0)\ C(b0\ \overline l0,l_{1}0)
\biggr]
\end{eqnarray}

Similar steps or using judicious replacements lead to
\begin{eqnarray}
\label{93}
\nonumber
M_{d}^{J,3}&=&{\delta (q'-q_{2})\over q'^{2}}\ 
\delta_{\lambda'\lambda_{2}}\ \delta_{I'I_{2}}\\
\nonumber
&\times&
(-)^{I'+J_{2}+l'+l_{2}+1}\ \sqrt{\hat j'\hat j_{2}\hat s'\hat s_{2}\hat J'}\
C(1-\!\!\mu\ J'M',J_{2}M_{2})\ 
\left\{ \matrix{1&j_{2}&j'\cr I'&J'&J_{2}}\right\}\\
\nonumber
&\times&
\biggl[ \delta_{l'l_{2}}\ i\ 4\pi\ \sqrt 6\ (-)^{l'+s'}\
\left\{ \matrix{l'&s'&j'\cr 1&j_{2}&s_{2}}\right\}\
\left\{ \matrix{1&1&1\cr 
        {1\over 2}&{1\over 2}&s_{2}\cr
        {1\over 2}&{1\over 2}&s'}\right \}\
\tilde H_{l'}(p'p_{2})\\
\nonumber
&+&i\ 240\pi\ \sqrt6\ (-)^{j_{2}}\
\sum_{i_{1}} (-)^{i_{1}}\ \hat i_{1}\
\left\{ \matrix{2&i_{1}&1\cr 1&1&1}\right\}\
\left\{ \matrix{2&i_{1}&1\cr l_{2}&s_{2}&j_{2}\cr l'&s'&j'} \right\} \
\left\{ \matrix{1&1&i_{1}\cr 
        {1\over 2}&{1\over 2}&s_{2}\cr
        {1\over 2}&{1\over 2}&s'}\right \}\\
&\times&
\sum_{a+b=2} {p'^{a}\ p_{2}^{b}\over \sqrt{(2a)!\ (2b)!}}\
\sum_{\overline l} \hat {\overline l}\ H_{\overline l}(p'p_{2})\
\left\{ \matrix{a&b&2\cr l_{2}&l'&\overline l}\right\} \
C(a0\ \overline l0,l'0)\ C(b0\ \overline l0,l_{2}0)
\biggr]
\end{eqnarray}

The last step is to perform the sum over $\mu$, eq.~(\ref{80}). Because in
eq.~(\ref{80}) $M_{d}$ is used together with $M_{b}$, we have to define
$\overline M_{d}$ consistently with $\overline M_{b}$ (eq.~(\ref{75})):
\begin{eqnarray}
\label{94}
\nonumber
M_{d}^{J,2}&\equiv& 
C(1\mu\ J_{1}M_{1},JM)\ (-)^{J_{1}-J}\ \overline M_{d}^{J,2}\\
M_{d}^{J,3}&\equiv& 
C(1-\!\!\mu\ J'M',J_{2}M_{2})\ \sqrt{\hat J'\over \hat J_{2}}\ 
\overline M_{d}^{J,3}
\end{eqnarray}
and we have the relation
\begin{equation}
\label{96}
\overline M_{d}^{J,2} (p\alpha ,p_{1}\alpha_{1})=
(-)^{l+l_{1}}\ \overline M_{d}^{J,3}(p_{1}\alpha_{1},p\alpha )
\end{equation}
Then the summation over $\mu$ can be done analytically using again
eq.~(\ref{74}). 

This concludes the presentation of 
the new PWD for all four terms of the TM 3NF. In this new approach one avoids
the pitfall of the old PWD and one can evaluate now the 3NF for
higher partial waves.

\section{Numerical Examples} 
\label{comp}

This section has two aims. The first is to show that the old and new PWD give
the same results for partial waves where the old PWD is still numerically
valid. The 
second is to show that the extension of the new PWD to higher partial waves
gives correct results.
Lower and higher partial waves refer to the outer states of type 1 in the
matrix elements of the 3NF. The number of intermediate states of type 2 or 3
in 
eqs.(\ref{29e}) and (\ref{29a}) are principally unlimited and the necessary
number of states for an accurate representation of the 3NF matrix elements
have to be determined numerically. We find, that in  eqs.(\ref{29e}) and
(\ref{29a}) the $\alpha''$ sums require two-body angular momenta up to
$j_{max}=5$. As already mentioned the additional sums in eq.~(\ref{29a}) over
$\alpha_{1J}$ and $\alpha_{2J}$ are essentially trivial, since the
corresponding quantum numbers are only marginally changed by the action of
$w_{4}^{(1)}|_{2,3}$ from the ones of the states of type 2 and 3.

We calculate the 3NF at 16 $p$-points below 10 fm$^{-1}$ and 16 $q$-points
below 20 fm$^{-1}$. According to our experience this is enough to describe
the 3NF \cite{HWG93} \cite{PhD}. The inner basis states of type 2 and 3
require 20 $p$- and $q$-points in order to achieve a fully converged
result. The $p$-range is extended to 16 fm$^{-1}$ in order to avoid
extrapolation. 

As mentioned in section \ref{old} the old PWD fails for $l\ge 4$ and
$\lambda\ge 4$. Therefore the largest two-body angular momentum, for which the
old PWD works, is $j=2$. Under this limitation we can compare the old and new
PWD. We choose $\Lambda=5.8\ m_{\pi}$ in eq.~(\ref{25}) and take as 
NN force the CD Bonn (np) \cite{Bonn}
force restricted to $j\le 2$, too.

\begin{figure}
\input{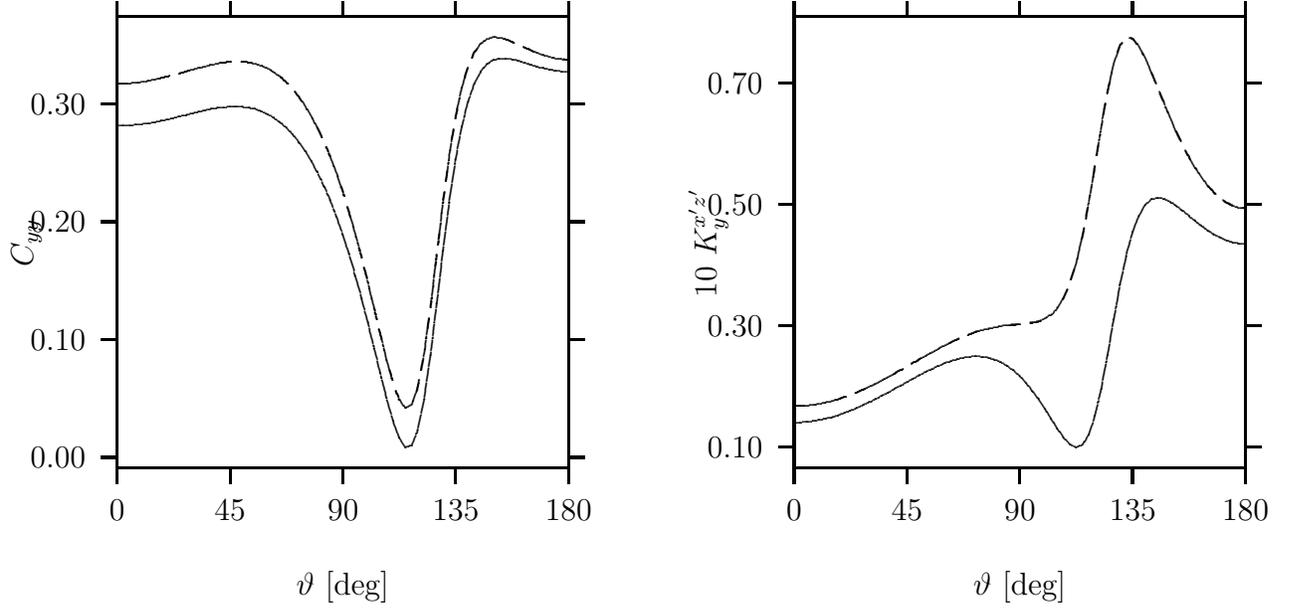}
\caption{The spin correlation coefficient $C_{yy}$ and the vector to tensor
  spin transfer coefficient $K_{y}^{x'z'}$ for elastic scattering at
  $E_{lab}=10.3$ MeV. The solid line is the result without 3NF, the
  overlapping long and
  short dashed lines are the results with 3NF using the old and new
  PWD, respectively.}
\label{elastic}
\end{figure}

In Fig.~\ref{elastic} we show two polarization observables, which show a big
effect of the 3NF. (The differential cross 
section is only marginally affected by the 3NF.) Fig.~\ref{elastic}
demonstrates that the predictions for the old and new PWD do completely
overlap. This 
is true for all other elastic observables, too.

\begin{figure}
\input{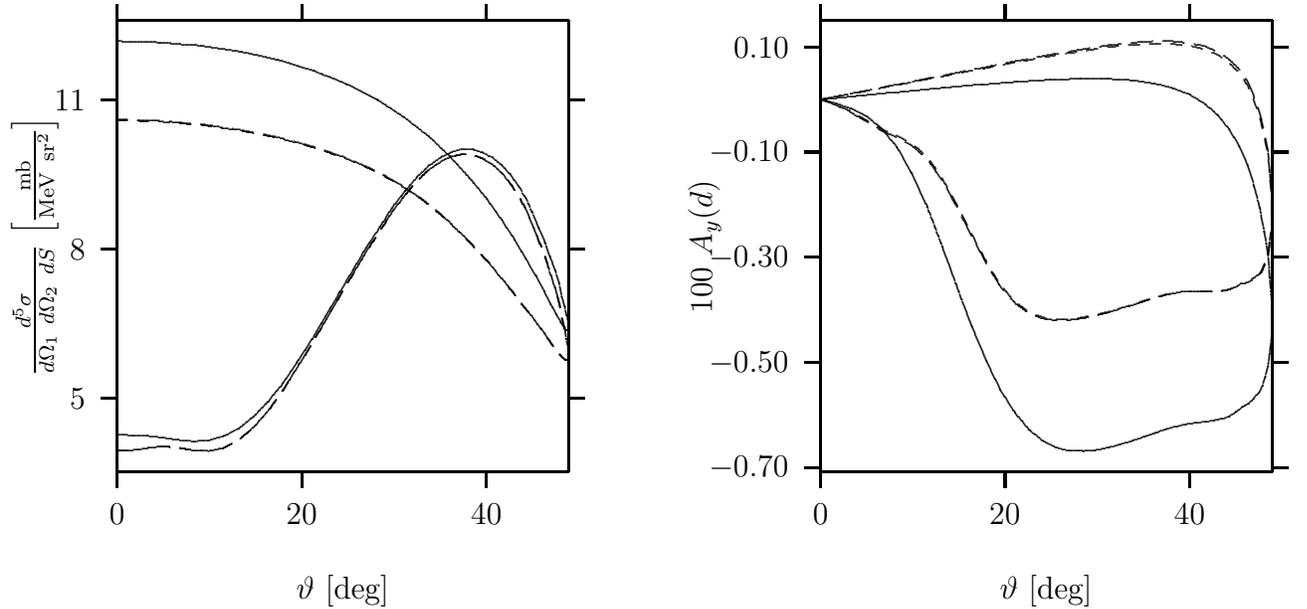}
\caption{The differential cross section and the deuteron analyzing power
  $A_{y}$ under np QFS conditions for the breakup process at  $E_{lab}=10.3$
  MeV as functions of the scattering angle of one nucleon. Description
  as in Fig.~3.}
\label{breakup}
\end{figure}

For the breakup process we calculated cross sections and all analyzing powers,
spin correlation coefficients and vector spin transfer coefficients
for standard kinematics. Again we chose as examples in Fig.~\ref{breakup}
observables with big 3NF effects. They are both observables under np QFS
conditions. Again the agreement of the two curves for
the old and new PWD is very good. This is generally true.
If one chooses $j_{max}=4$ for the inner basis states
2 and 3 instead of $j_{max}=5$ the deviations in the observables to the full
result are up to a few percent.

In order to show that also the extension of the new PWD to higher partial
waves works well we performed a calculation of the triton binding energy with
$j_{max}=4$. As NN force we used here the AV14 potential \cite{AV14}. Our
result for the triton binding energy is $-9.36$ MeV, which agrees exactly with
the result of the Los Alamos - Iowa group \cite{LA}. 

We obtained the results presented up to now
not by using eq.~(\ref{29e}) but the form
\begin{eqnarray}
\label{96a}
\nonumber
&&\Bra 1{pq\alpha }V_{4}^{(1)}\ (1+P)\ket{p'q'\alpha'}_{1}\\
&=&\intsum''\
\Bra 1{pq\alpha }V_{4}^{(1)}\ket{p''q''\alpha''}_{1}\
\Bra 1{p''q''\alpha''}(1+P)\ket{p'q'\alpha'}_{1}
\end{eqnarray}
This is identical to eq.~(\ref{29e}) if the $\alpha''$ summation is
unlimited. In the 
calculations, however, we restricted the double primed channels to the
physical channels, which is an approximation. The old code for the 3NF did
anyhow not allow the higher partial waves. Now using eq.~(\ref{29e}) and the
new PWD we can check the quality of that approximation. We compare the results
for triton binding energies based on the new PWD of the 3NF achieved via
eq.~(\ref{29e}) (denoted by $E_{b}^{(1)}$) with the fully converged
$\alpha''$-sum to the result gained through eq.~(\ref{96a}) (denoted by
$E_{b}^{(2)}$), where the
$\alpha''$-sum is restricted to the same number of partial waves as used in
the outer states. The latter number is fixed by choosing a certain maximal
total two-body angular momentum $j_{max}$. Our results are shown in table
\ref{tab1}. For $j_{max}=5$ we achieved convergence, since the two binding
energies $E_{b}^{(1)}$ and $E_{b}^{(2)}$ are equal. The effect of the
truncation is large for $j_{max}=2$, whereas $E_{b}^{(1)}$ under the same
restriction of outer partial waves is already close to the final result. We
also see, that the $j_{max}=4$ result for $E_{b}^{(2)}$ is close to the
correct number $E_{b}^{(1)}$.

The results with the NN force only are nearly converged at $j_{max}=3$, whereas
the interplay with the 3NF causes a decrease in binding energy for $j_{max}=3$
and only for $j_{max}=4$ the final result is essentially reached.

\section{Summary and Outlook}
\label{sum}

We demonstrated in section \ref{old} that the way used up to now to  
decompose 3NF's in momentum space into partial waves \cite{CG81} leads
unavoidably to numerical instabilities. They occur necessarily if the momenta
and the angular momentum quantum numbers are high. The applications carried
through up to now are not affected by that numerical instability, since the
3NF has been applied up to now only in low partial waves. For higher energies,
however, high angular momenta are activated and the old PWD cannot be used.

In section \ref{new} we presented a
new PWD. The basic idea is to treat the 3NF in the example of the
$2\pi$-exchange as if it would be a sequence of two pion exchanges among
different pairs. This requires the choice of different Jacobi variables
adopted to the two different two-body subsystems. Consequently in between
there has to occur a recoupling of the Jacobi variables. In this manner we not
only achieve a much simpler analytical form of the partial wave decomposed 3NF,
but also an expression, which is free of any artificial singularity and
numerically stable in all angular momentum states. The recoupling between the
two meson exchanges and also to the external basis states, which require the
third possibility for the choice of a two-body subsystem, are performed with
the aid of certain permutation operators, which are laid out in appendix
\ref{perm}. The application of these permutation operators lead to
two-dimensional interpolations, which require very efficient and accurate
tools. To that purpose we have rewritten standard hermitian cubic splines into
a new form, which is shown in appendix \ref{spline}.

Finally we have shown in section \ref{comp} two numerical examples for the
reliability of the new PWD of the 3NF. We used the TM 3NF model and calculated
in the first example observables in
elastic nd scattering and the breakup process at $E_{lab}=10.3$ MeV. The
forces were restricted to $j_{max}=2$ so that the old PWD could also be
used. For 
all observables we found a very good agreement between the predictions 
achieved with the old and new PWD. In the second example we
extended the new PWD of the TM 3NF to higher partial waves with $j_{max}=5$
and calculated 
the triton binding energy. We could confirm the result of the Los Alamos -
Iowa group truncated at $j_{max}=4$. The convergence of the triton binding
energies as a function of $j_{max}$ as displayed in table \ref{tab1} shows,
that in contrast to the case of NN forces only, the interplay with the 3NF can
lead to a more oscillating approach of the limiting value.
More benchmark calculations are
underway. 

Summarizing we can say that we are now able to calculate the 3NF up to any
partial wave, limited only by computer resources. This new scheme can of course
be taken over from the $2\pi$-exchange, exemplified here, to any other
two-meson exchange.

It is interesting to note that the new PWD has still another application. If
one wants to calculate 
three-body meson exchange currents as they occur for example in electron
scattering on $^{3}$He ($^{3}$H) or heavier nuclei one faces exactly the same
problems as described here. In order to get correct results for all partial
waves one has to follow for the PWD of the three-body meson exchange currents
the scheme which we have introduced above.

\begin{center}
{\bf Acknowledgements}
\end{center}

This work was supported by the Deutsche Forschungsgemeinschaft, the 
Polish Committee for Scientific Research under Grant No. 2 P302 104 06 and 
the Polish-American Maria Sk{\l}odowska-Curie II FUND under Grant No. 
MEN/NSF-94-161. The numerical calculations have been performed on the 
CRAY T90 and the CRAY T3E of the H\"ochstleistungsrechenzentrum in J\"ulich,
Germany.  

\begin{appendix}

\section{The Permutation Operators}
\label{perm}

In the Faddeev scheme \cite{HWG93} \cite{Buch} we apply the permutation
operator 
$P=P_{12}P_{23}+P_{13}P_{23}$ always to states which are
anti-symmetric in the 
two-body subcluster 2-3. Regarding eqs.~(\ref{29a}) and (\ref{96a}) we see the
recoupling matrix elements $\Scal 1{pq\alpha }{p'q'\alpha'}_{2}$, 
$\Scal 2{pq\alpha_{J} }{p'q'\alpha_{J}'}_{3}$, and
$\Scal 3{pq\alpha }{p'q'\alpha'}_{1}$. Since $\ket{}_{3}\equiv
P_{13}P_{23}\ket{}_{1}$ and $\ket{}_{2}\equiv P_{12}P_{23}\ket{}_{1}$ one
easily finds that all three matrix elements are of the type $\Bra
1{}P_{12}P_{23}\ket{}_{1}$. The expressions given in \cite{HWG93} and
\cite{report} for $P$ are in fact evaluated there with the help of $\Bra
1{}P_{12}P_{23}\ket{}_{1}$ and an originally occurring phase factor
$(-)^{l'+s'+t'}$ has been replaced by $(-1)$, since there only antisymmetric
two-body states have been used. Redoing that replacement one finds the desired
matrix element from the one of $P$ given in \cite{HWG93} and \cite{report} by
multiplying it with ${1\over 2}(-)^{l'+s'+t'+1}$. For the states without
iso-spin one has just to drop the iso-spin factor.

Out of the many various forms to evaluate permutation operators \cite{HKWG94}
we need here three of them. The recoupling between the quasi two-body force
quantities requires 
\begin{equation}
\label{101}
\Bra 1{pq\alpha_{J}}P_{12}P_{23}\ket{p'q'\alpha_{J}'}_{1}
={1\over 2}\ (-)^{l'+s'+1}\
\int_{-1}^{+1}dx\ {\delta(p-\pi)\over p^{l+2}}\ {\delta(p'-\pi')\over
  p'^{l'+2}}\ 
G_{\alpha_{J}\alpha_{J}'}(qq'x)
\end{equation}
with
\begin{eqnarray}
\label{102}
\nonumber
\pi&=&\sqrt{{1\over 4}q^{2}+q'^{2}+qq'x}\\
\pi'&=&\sqrt{q^{2}+{1\over 4}q'^{2}+qq'x}
\end{eqnarray}
and
\begin{equation}
\label{103}
G_{\alpha_{J}\alpha_{J}'}(qq'x)
=\sum_{k} P_{k}(x) \sum_{l_{1}+l_{2}=l} \sum_{l'_{1}+l'_{2}=l'}
q^{l_{2}+l_{2}'}\ q'^{l_{1}+l_{1}'}\
g_{\alpha_{J}\alpha_{J}'}^{kl_{1}l_{2}l'_{1}l'_{2}}
\end{equation}
The purely geometrical quantity
$g_{\alpha_{J}\alpha_{J}'}^{kl_{1}l_{2}l'_{1}l'_{2}}$ has to be taken now
without iso-spin:
\begin{eqnarray}
\label{104}
\nonumber
  g_{\alpha_{J} \alpha'_{J}}^{kl_1l_2l_1'l_2'}
  &=&-\sqrt{\hat l \ \hat s \ \hat j \ \hat {\lambda} \ \hat I
           \ \hat l' \ \hat s' \ \hat j' \ \hat {\lambda '} \
           \hat I' }\\
  &\times&\sum_{LS} \hat L \hat S \ 
\left\{ \matrix{ {1\over 2}&{1\over 2}&s\cr {1\over 2}&S&{s'}}\right\} \
\left\{ \matrix{l&s&j\cr{\lambda}&{1\over 2}&I\cr L&S&J}\right\} \
\left\{ \matrix{{l'}&{s'}&{j'}\cr {\lambda '}&{1\over 2}&{I'}\cr L&S&J}
\right\} \nonumber \\
  &\times&\hat k \ \left( {1\over 2} \right)^{l_2+l_1'}\
          \sqrt{{(2l+1)!\over (2l_1)!\ (2l_2)!}}\
          \sqrt{{(2l'+1)!\over (2l_1')!\ (2l_2')!}}\nonumber \\
  &\times&\sum_{ff'} 
\left\{ \matrix{{l_1}&{l_2}&l\cr {\lambda}&L&f}\right\} \
\left\{ \matrix{{l_2'}&{l_1'}&{l'}\cr {\lambda '}&L&{f'}}\right\} \
C({l_2}0\ {\lambda}0,f0)\ C({l_1'}0\ {\lambda '}0,{f'}0)\nonumber \\
  &\times&
\left\{ \matrix{f&{l_1}&L\cr {f'}&{l_2'}&k}\right\} \
C(k0\ {l_1}0,{f'}0)\ C(k0\ {l_2'}0,f0)
\end{eqnarray}

The very left recoupling coefficient has to be taken in the form
where the $\delta$-functions act both to the right:
\begin{equation}
\label{105}
\Bra 1{pq\alpha}P_{12}P_{23}\ket{p'q'\alpha'}_{1}
={1\over 2}\ (-)^{l'+s'+t'+1}\
\int_{-1}^{+1}dx\ {\delta(p'-\tilde \pi)\over p'^{l'+2}}\ 
{\delta(q'-\tilde \chi)\over q'^{\lambda'+2}}\
\tilde G_{\alpha\alpha'}(pqx)
\end{equation}
with
\begin{eqnarray}
\label{106}
\nonumber
\tilde \pi&=&\sqrt{{1\over 4}p^{2}+{9\over 16}q^{2}+{3\over 4}pqx}\\
\tilde \chi&=&\sqrt{p^{2}+{1\over 4}q^{2}-pqx}
\end{eqnarray}
and $\tilde G_{\alpha\alpha'}(pqx)$ can be taken for instance from
\cite{report}. 

Finally the recoupling coefficient from the states of type 1 to 3 require a
form where the
$\delta$-functions act both to the left:
\begin{equation}
\label{107}
\bra{pq\alpha}P_{12}P_{23}\ket{p'q'\alpha'}
={1\over 2}\ (-)^{l'+s'+t'+1}\
\int_{-1}^{+1}dx\ {\delta(p-\tilde {\tilde \pi})\over p^{l+2}}\ 
{\delta(q-\tilde {\tilde \chi})\over q^{\lambda+2}}\
\tilde G_{\alpha'\alpha}(p'q'x)
\end{equation}
with
\begin{eqnarray}
\label{108}
\nonumber
\tilde {\tilde \pi}
&=&\sqrt{{1\over 4}p'^{2}+{9\over 16}q'^{2}-{3\over 4}p'q'x}\\
\tilde {\tilde \chi}&=&\sqrt{p'^{2}+{1\over 4}q'^{2}+p'q'x}
\end{eqnarray}
In eq.~(\ref{107}) we need the same quantity $\tilde G$ as in eq.~(\ref{105}),
but with interchanged arguments.

\section{Cubic Hermitian Splines}
\label{spline}

In appendix \ref{perm} we encountered two-fold interpolations. They have to be
performed for very many
channels and should be as optimal as possible.
We found that our usual
basis splines \cite{GHN82} are not efficient enough to perform these two-fold
interpolations within a reasonable time. Therefore we need a more efficient
interpolation algorithm.

Such an algorithm can be found by using cubic hermitian splines. Although
cubic hermitian splines are well known in the literature (see for example
\cite{SPA90}), we will give here a short introduction in order to explain our
way to use them.

Consider a function $f(x)$ given at certain grid points ${\rm x}_{i}$,
$i=1,\ldots ,n$. Let $x$ be
positioned in the interval $[{\rm x}_{i},{\rm x}_{i+1}]$. For the sake of
simpler notation we call the end points ${\rm x}_{i}\equiv x_{1}$ and  ${\rm
  x}_{i+1}\equiv x_{2}$. 
Then one defines an unique cubic
polynomial $f_{i}(x)$ by the following constraints:
\begin{eqnarray}
\label{109}
\nonumber
f_{i}(x_{1})&=&f(x_{1})\\
\nonumber
f_{i}(x_{2})&=&f(x_{2})\\
\nonumber
f'_{i}(x_{1})&=&f'(x_{1})\\
f'_{i}(x_{2})&=&f'(x_{2})
\end{eqnarray}
Therefore these interpolating functions $f_{i}(x)$ and their derivatives
$f'_{i}(x)$ are
continuous at the grid points ${\rm x}_{i}$.
They are given by
\begin{equation}
\label{110}
f_{i}(x)=f(x_{1})\phi_{1}(x)+f(x_{2})\phi_{2}(x)
        +f'(x_{1})\phi_{3}(x)+f'(x_{2})\phi_{4}(x)
\end{equation}
in terms of the spline functions
\begin{eqnarray}
\label{111}
\nonumber
\phi_{1}(x)&=&
\left( {(x_{2}-x)^{2}\over (x_{2}-x_{1})^{3}}
\left( (x_{2}-x_{1})+2(x-x_{1})\right)
\right)\\
\nonumber
\phi_{2}(x)&=&
\left( {(x_{1}-x)^{2}\over (x_{2}-x_{1})^{3}}
\left( (x_{2}-x_{1})+2(x_{2}-x)\right)
\right)\\
\nonumber
\phi_{3}(x)&=&
{(x-x_{1})(x_{2}-x)^{2}\over (x_{2}-x_{1})^{2}}\\
\phi_{4}(x)&=&
{(x-x_{1})^{2}(x-x_{2})\over (x_{2}-x_{1})^{2}}
\end{eqnarray}

We approximate the derivatives $f'(x_{1})$ and $f'(x_{2})$ with the help
of a quadratic
polynomial, which is uniquely defined by the function values at the grid point
and its two neighbours. Calling ${\rm x}_{i-1}=x_{0}$ and
${\rm x}_{i+1}=x_{3}$ we get
\begin{eqnarray}
\label{112}
\nonumber
f'(x_{1})&\approx& 
f(x_{2})\ {x_{1}-x_{0}\over x_{2}-x_{1}}\ {1\over x_{2}-x_{0}}
-f(x_{0})\ {x_{2}-x_{1}\over x_{1}-x_{0}}\ {1\over x_{2}-x_{0}}\\
\nonumber
&+&f(x_{1})\ 
\left( {x_{2}-x_{1}\over x_{1}-x_{0}}-{x_{1}-x_{0}\over x_{2}-x_{1}}
\right)\ {1\over x_{2}-x_{0}}\\
\nonumber
f'(x_{2})&\approx& 
f(x_{3})\ {x_{2}-x_{1}\over x_{3}-x_{2}}\ {1\over x_{3}-x_{1}}
-f(x_{1})\ {x_{3}-x_{2}\over x_{2}-x_{1}}\ {1\over x_{3}-x_{1}}\\
&+&f(x_{2})\ 
\left( {x_{3}-x_{2}\over x_{2}-x_{1}}-{x_{2}-x_{1}\over x_{3}-x_{2}}
\right)\ {1\over x_{3}-x_{1}}
\end{eqnarray}
At the end points $x_{1}$ and $x_{n}$
we define the
quadratic polynomial by $f({\rm x}_{1})$, $f({\rm x}_{2})$ and $f({\rm
  x}_{3})$ and by $f({\rm x}_{n-2})$, $f({\rm x}_{n-1})$ and $f({\rm x}_{n})$,
respectively. This is achieved by putting
$x_{0}={\rm x}_{3}$ in the
first case and $x_{3}={\rm x}_{n-2}$ in the second case.

Insertion of eq.~(\ref{112}) into eq.~(\ref{110}) yields
\begin{equation}
\label{113}
f_{i}(x)=\sum_{j=0}^{3} S_{j}(x)\ f(x_{j})
\end{equation}
for the interpolating function in the $i$-th interval of the grid
points. Thereby we are lead to the  
modified spline functions
\begin{eqnarray}
\label{114}
\nonumber
S_{0}(x)&=&-\phi_{3}(x)\ {x_{2}-x_{1}\over x_{1}-x_{0}}\ {1\over x_{2}-x_{0}}\\
\nonumber
S_{1}(x)&=&\phi_{1}(x)+\phi_{3}(x)\ 
\left( {x_{2}-x_{1}\over x_{1}-x_{0}}-{x_{1}-x_{0}\over x_{2}-x_{1}}
\right)\ {1\over x_{2}-x_{0}}
-\phi_{4}(x)\ {x_{3}-x_{2}\over x_{2}-x_{1}}\ {1\over x_{3}-x_{1}}\\
\nonumber
S_{2}(x)&=&\phi_{2}(x)+\phi_{3}(x)\
{x_{1}-x_{0}\over x_{2}-x_{1}}\ {1\over x_{2}-x_{0}}
+\phi_{4}(x)\ 
\left( {x_{3}-x_{2}\over x_{2}-x_{1}}-{x_{2}-x_{1}\over x_{3}-x_{2}}
\right)\ {1\over x_{3}-x_{1}}\\
S_{3}(x)&=&\phi_{4}(x)\ {x_{2}-x_{1}\over x_{3}-x_{2}}\ {1\over x_{3}-x_{1}}
\end{eqnarray}
Eq.~(\ref{113}) is very well suited for the numerical usage. The
modified spline functions $S_{j}(x)$ are independent of the function values
and depend only on the grid points and the actual position $x$.
Therefore they can be prepared
beforehand. This is very important if one has to interpolate very many
functions given at the same grid points as we have to do in our 3NF code.

The form of eq.~(\ref{113}) is the same as the one found in \cite{GHN82} for
basis splines. The difference is, that the sum for the basis splines runs over
the whole grid, whereas the sum for the hermitian splines in eq.~(\ref{113})
has only four terms related to the four grid points nearest to
the interpolation point $x$. (Basis splines are global splines, whereas
hermitian splines are local.) Assuming a grid of typically 20 points the
one-dimensional 
interpolation using hermitian splines needs only $1\over 5$ operations
compared to basis splines. For a two-dimensional interpolation the number of
operations is reduced by a factor of $1\over 25$.

For the two-dimensional interpolation one has to make a bi-cubic ansatz for
the interpolating functions $f_{ij}(x,y)$. To define a bi-cubic function
uniquely we need 16 constrains, which we choose as
\begin{eqnarray}
\label{115}
\nonumber
f_{ij}(x_{1},y_{1})&=&f(x_{1},y_{1})\\
\nonumber
\left .{\partial f_{ij}(x,y_{1})\over \partial x}\right |_{x=x_{1}}
&=&\left .{\partial f(x,y_{1})\over \partial x}\right |_{x=x_{1}}\\
\nonumber
\left .{\partial f_{ij}(x_{1},y)\over \partial y}\right |_{y=y_{1}}
&=&\left .{\partial f(x_{1},y)\over \partial y}\right |_{y=y_{1}}\\
\left .{\partial ^{2}f_{ij}(x,y)\over \partial x\ \partial y}
\right |_{x=x_{1},y=y_{1}}
&=&\left .{\partial ^{2}f(x,y)\over \partial x\ \partial y}
\right |_{x=x_{1},y=y_{1}}
\end{eqnarray}
and identical expressions for the other three points $(x_{1},y_{2})$,
$(x_{2},y_{1})$, and $(x_{2},y_{2})$, respectively. Hereby the four grid
points 
$(x_{1},y_{1})$, $(x_{1},y_{2})$,
$(x_{2},y_{1})$ and $(x_{2},y_{2})$ are nearest neighbours to the
interpolation point $(x,y)$ in the $xy$-plane.

The partial derivatives are approximated as in the one-dimensional
case. The second derivative is estimated by
a bi-quadratic polynomial, which is
uniquely given by the function value at the specific grid point
and the function values of the eight surrounding grid points.

Following these steps one yields
\begin{equation}
\label{116}
f_{ij}(x,y)=\sum_{k=0}^{3}\sum_{l=0}^{3} S^{(2)}_{kl}(x,y)\ f(x_{k},y_{l})
\end{equation}
for the interpolating function. It is a fairly easy exercise to show that the
two-dimensional spline functions $S^{(2)}_{kl}(x,y)$ are simply given by
\begin{equation}
\label{117}
S^{(2)}_{kl}(x,y)=S_{k}(x)\ S_{l}(y)
\end{equation}
Analogical equations hold for interpolations in more than two dimensions.

To our experience one- and two-dimensional interpolations using hermitian
splines are at least of the same accuracy as interpolations based on basis
splines.

\end{appendix}

\begin{table}
\begin{tabular} {c|c|c|c}
$j_{max}$&$E_{b}^{NN}$&$E_{b}^{(1)}$&$E_{b}^{(2)}$\\
\hline
2&7.58&9.34&9.51\\
3&7.67&9.27&9.25\\
4&7.68&9.34&9.36\\
5&7.69&9.32(9)&9.32(4)
\end{tabular}
\caption{\label{tab1} Triton binding energies calculated using the AV14 NN
  force alone ($E_{b}^{NN}$) and together with the TM 3NF
  ($\Lambda=5.8m_{\pi}$). For explanation of 
  $E_{b}^{(1)}$ and $E_{b}^{(2)}$ see text.}
\end{table}

\input{prepictex.tex}
\input{pictex.tex}
\input{postpictex.tex}
\font\fiverm=cmr5
\font\sevenrm=cmr7

\end{document}